\documentclass[aps,superscriptaddress,prb,amsmath,amssymb,reprint,longbibliography]{revtex4-2}

\usepackage[USenglish]{babel}
\usepackage[T1]{fontenc}
\usepackage[utf8]{inputenc}
\usepackage{graphicx}
\usepackage{upgreek}
\usepackage[unicode]{hyperref}
\usepackage{multirow}
\usepackage{verbatim} 
\usepackage[usenames, dvipsnames]{color} 
\usepackage{verbatim} 
\usepackage[usenames, dvipsnames]{color} 
\usepackage{soul}
\usepackage{fixltx2e}
\usepackage{booktabs} 
\usepackage{siunitx}  
\usepackage{titlesec}

\titlespacing*{\section}{0pt}{12pt}{6pt} 
\begin{document}


\setlength{\bibsep}{6pt} 
	
\title{Layer-Dependent Spin Properties of Charge Carriers in\\ Vertically Coupled Telecom Quantum Dots}

\date{\today}

\author{Marius~Cizauskas}
\email{email: marius.cizauskas@tu-dortmund.de}
\affiliation{Experimentelle Physik 2, Technische Universit\"at Dortmund, 44221 Dortmund, Germany}
\affiliation{Department of Physics and Astronomy, University of Sheffield, Sheffield, UK}

\author{A.~Kors}
\affiliation{Institute of Nanostructure Technologies and Analytics (INA), CINSaT, University of Kassel, D-34132 Kassel, Germany}

\author{J.~P.~Reithmaier}
\affiliation{Institute of Nanostructure Technologies and Analytics (INA), CINSaT, University of Kassel, D-34132 Kassel, Germany}

\author{A.~Mark~Fox}
\affiliation{Department of Physics and Astronomy, University of Sheffield, Sheffield, UK}

\author{M.~Benyoucef}
\email{email: m.benyoucef@physik.uni-kassel.de}
\affiliation{Institute of Nanostructure Technologies and Analytics (INA), CINSaT, University of Kassel, D-34132 Kassel, Germany}

\author{Manfred~Bayer}
\affiliation{Experimentelle Physik 2, Technische Universit\"at Dortmund, 44221 Dortmund, Germany}

\author{Alex~Greilich}
\email{email: alex.greilich@tu-dortmund.de}
\affiliation{Experimentelle Physik 2, Technische Universit\"at Dortmund, 44221 Dortmund, Germany}

\begin{abstract}
We investigate the spin properties of charge carriers in vertically coupled InAs/InAlGaAs quantum dots grown by molecular beam epitaxy, emitting at telecom C-band wavelengths, with a silicon $\delta$-doped layer. Using time-resolved pump-probe Faraday ellipticity measurements, we systematically study single-, two-, and four-layer quantum dot (QD) configurations to quantify how vertical coupling affects key spin-coherence parameters. Our measurements reveal distinct layer-dependent effects: (1) Adding a second QD layer flips the resident charge from electrons to holes, consistent with optically induced electron tunneling into lower-energy dots and resultant hole charging. (2) Starting from the four-layer sample, the pump-probe signal develops an additional non-oscillating, decaying component absent in single- and two-layer samples, attributed to multiple layer growth changing the strain environment, which reduces heavy-hole and light-hole mixing. (3) With four-layers or more, hole spin mode locking (SML) can be observed, enabling quantitative extraction of the hole coherence time $T_2 \approx 13$\,ns from SML amplitude saturation. We also extract longitudinal spin relaxation ($T_1$) and transverse ($T_2^*$) spin dephasing times, g-factors, and inhomogeneous dephasing parameters for both electrons and holes across all layer configurations. The hole spin dephasing times $T_2^*$ remain relatively constant (2.26-2.73\,ns) across layer counts, while longitudinal relaxation times $T_1$ decrease with increasing layers (from 1.03\,$\mu$s for single-layer to 0.31\,$\mu$s for four-layer samples). These findings provide potential design guidelines for engineering spin coherence in telecom-band QDs for quantum information applications.
\end{abstract}

\maketitle

\section{Introduction}

The advancement of quantum information and communication technologies relies on the development of high-quality quantum systems that can maintain coherence over extended timescales while operating under practical conditions. Semiconductor quantum dots (QDs) have emerged as versatile platforms for quantum information applications due to their potential for optical control of spin states, scalable fabrication, possibility of on-chip integration and use as high purity single-photon sources~\cite{Vajner_2022, Miyazawa_2016, Pawel_2021}.

For quantum information applications, operation at telecom wavelengths (1.3-1.6 \si{\micro\meter}) offers practical advantages, which is the possibility to use existing telecom fiber infrastructure. Efficient QD coupling to single-mode fiber has been demonstrated with additional nanostructures designed to reduce losses~\cite{Bauer_2021, Rahaman_2024}. InAs/InGaAs/InP QDs are a promising solution for telecom-band quantum information applications, as they can be engineered to emit efficiently in the C-band while maintaining the spin coherence properties necessary for quantum applications~\cite{Mikhailov_2018, Evers_2024}.

The ultimate performance of QD spin-based quantum information systems is governed by the coherence properties of the carrier spins, which are subject to decoherence from hyperfine interactions, phonon scattering, and structural asymmetries~\cite{Mikhailov_2018, Smirnov_2020}. Molecular beam epitaxy (MBE) has been the predominant growth technique for high-quality QD, offering precise control over layer thickness, composition, and doping profiles~\cite{Li_2023}. However, the specific growth conditions and structural parameters can significantly influence the resulting spin coherence properties.

An important consideration in QD design is the impact of layer structure on spin dynamics. While single-layer QD structures offer simplicity, multi-layer configurations can provide enhanced optical signals and the potential for engineering inter-layer interactions~\cite{Stinaff2006,Kim2008,wigger_2023}. The samples investigated in this work are InAs/InAlGaAs QDs grown by MBE, with a structure identical to that reported by Ref.~\cite{Evers_2024}, with the difference that in Ref.~\cite{Evers_2024} the sample consists of eight-layers. We focus on comparing the spin properties of single-, two-, and four-layer stacks of QD configurations to quantify how additional layers affect key spin-coherence parameters. Using time-resolved pump–probe Faraday ellipticity, we extract longitudinal and transverse relaxation times, $g$-factors, and inhomogeneous dephasing for electrons and holes.

\begin{figure*}[]
\centering
\includegraphics[width=11cm]{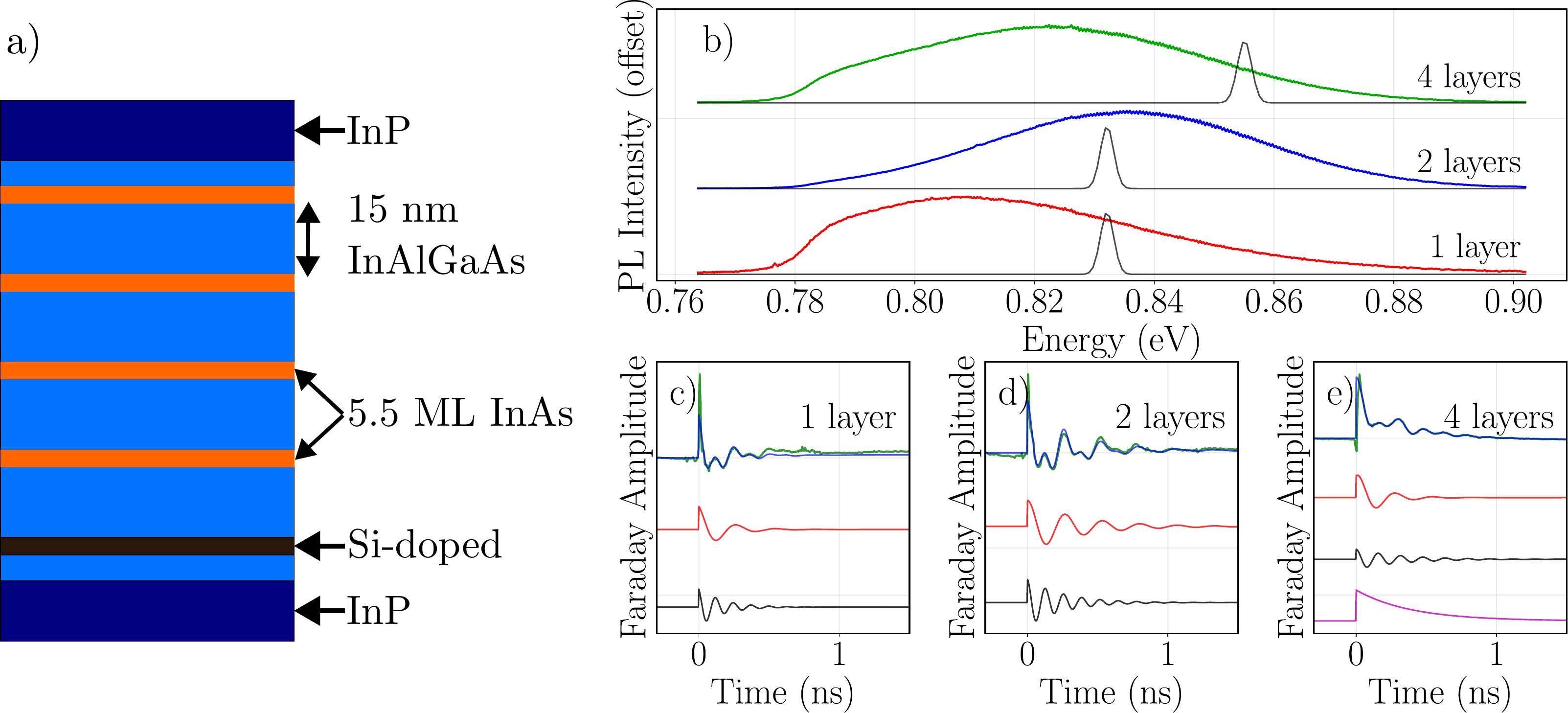}
\caption{(a) Example of a four-layer InAs/InAlGaAs stacked QD sample structure. Photoluminescence (PL) spectra and Faraday ellipticity measurements for QD samples with varying numbers of layers are presented in the subsequent plots. Panel (b) shows PL intensity spectra (offset for clarity) for samples with one layer (red), two layers (blue), and four layers (green), with the laser excitation spectrum shown as a gray line for each sample. All PL spectra were obtained with a 2.33\,eV laser as the excitation source. For Faraday ellipticity, 0.832 eV excitation was used for the single- and two-layer samples, and 0.855 eV for the four-layer sample. Panels (c), (d), and (e) show time-resolved Faraday ellipticity amplitude measurements for the 1-layer, 2-layer, and 4-layer samples respectively, where the red traces correspond to hole spin precession, the black traces correspond to electron spin precession, and the purple trace (in panel e) represents an additional exponential decay component that may arise from inter-dot coupling effects or other multi-dot interactions. The component measurements were all done at a transverse magnetic field strength of 0.4~T. The pump and probe powers for the single-layer sample were 15~mW and 2~mW, for the two-layer sample 20~mW and 2~mW, and for the four-layer sample 10~mW and 2~mW. All measurements were done at 6\,K.}
\label{fig:one}
\end{figure*}

Across otherwise-identical stacks, we uncover clear layer-dependent trends. First, adding a second QD layer flips the resident charge from electrons to holes, consistent with optically induced electron tunneling into lower-energy dots and resultant hole charging. Second, in four-layer stacks, the pump–probe signal develops an additional non-oscillating decay component absent in one- and two-layer samples, which we attribute to potential inter-layer coupling effects that modify the valence-band structure and reduce HH–LH mixing. Finally, the four-layer sample supports hole spin mode locking (SML), enabling the quantitative extraction of the hole coherence time $T_2 \approx 13$\,ns from SML amplitude saturation.

This layer-resolved study identifies how vertical coupling reshapes spin properties in telecom-band QDs and provides design guidelines for engineered coherence.


\section{Experimental details}
The QD samples investigated in this study are fabricated using MBE Stranski-Krastanov (SK) growth method on (100)-oriented InP substrates. The samples contain InAs QDs embedded within In\textsubscript{0.53}Al\textsubscript{0.24}Ga\textsubscript{0.23}As barrier layers, following the same structural design and growth parameters as reported by Ref.~\cite{Evers_2024}, see Fig.~\ref{fig:one}a. The only difference between our samples is the number of layers.

Each InAs layer consists of 5.5 monolayers (equivalent to approximately 1.65 nm) with QDs having a surface density of approximately 10\textsuperscript{10} cm\textsuperscript{-2}. A silicon $\delta$-doped layer is positioned 15 nm below the QD layers with a doping concentration of 10\textsuperscript{10} cm\textsuperscript{-2} to provide resident electrons within the QDs. In the two-layer sample, the QD layers are separated by 15 nm of In\textsubscript{0.53}Al\textsubscript{0.24}Ga\textsubscript{0.23}As barrier material. The quaternary barrier composition serves the dual purpose of providing enhanced carrier confinement.

Optical measurements are conducted using a liquid helium bath cryostat equipped with temperature control capabilities, allowing measurements across a temperature range from 6\,K to 60\,K. Magnetic field application is achieved through a superconducting magnet system capable of providing fields up to 4\,T. The magnetic field orientation can be configured either parallel to the optical excitation axis (Faraday configuration) or perpendicular to it (Voigt configuration) by mechanical rotation of the sample mount and cryostat.

The optical excitation source consists of a mode-locked Ti:Sapphire laser (MiraHP) operating at a repetition frequency of 76.7\,MHz, which pumps an optical parametric oscillator (APE OPO) to generate tunable pulses in the telecom wavelength range. The laser output provides pulses with approximately 2\,nm spectral bandwidth. The generated beam is divided into pump and probe paths using a beam splitter, with the probe beam directed through a computer-controlled mechanical delay stage for temporal scanning.

Both pump and probe beams are focused onto the sample surface, with the pump beam diameter set to 150\,\textmu m and the probe beam focused to 100\,\textmu m diameter to ensure complete spatial overlap. The excitation energy is tuned to coincide with the photoluminescence peak at approximately 0.83\,eV to achieve resonant excitation of the QD ensemble. 

Spin polarization in the QDs is generated using circularly polarized pump pulses, achieved through an electro-optic modulator that alternates the pump polarization between $\sigma^+$ and $\sigma^-$ states at frequencies ranging from 1\,kHz to 2\,MHz (with 10\,kHz used for Voigt geometry measurements). The probe beam is linearly polarized, and the intensity is modulated using a photoelastic modulator operating at 86\,kHz, followed by a Glan-Thompson polarizer.

Spin-dependent optical rotation of the probe beam is detected through Faraday ellipticity measurements~\cite{Yugova_2009}. Following interaction with the sample, the probe beam passed through a 10\,\textmu m diameter pinhole to minimize detection of scattered pump light. A quarter-wave plate converts the circular components of an elliptically polarized probe into linearly polarized components with orthogonal orientations. These components are separated using a Wollaston prism and directed onto balanced photodiodes for differential detection.

The photodiode signals are processed using lock-in amplification at the difference frequency between the pump and probe modulation frequencies, providing an enhanced signal-to-noise ratio and rejection of common-mode noise sources. This double-modulation detection scheme enables measurement of small spin-induced optical rotations with high sensitivity.


\section{Experimental Results}
Figure~\ref{fig:one}b shows the photoluminescence (PL) spectra of all the samples. The PL comparison of the different samples shows consistency; most of the change is seen in where the excitation of the sample is done. The pump-probe excitation energy was chosen by recording the electron and hole precession signal intensities at different wavelengths, while keeping the excitation power constant, and then selecting the wavelength with the largest signal amplitude. In the case of the 1- and 4-layer samples, the result is consistent and can be explained by the fact that at higher energies, more dots are excited, which results in higher signal amplitude. In the 4-layer sample, the excitation is shifted to higher energies to obtain higher amplitude for the hole spin component due to spin mode-locking effect discussed later in the paper. In the case of a 2-layer sample, the excitation is placed at a lower energy relative to the PL center. Since we have a doping layer close to the bottom layer, adding additional layers causes a change in charge distribution. This can cause the carrier precession amplitude peak wavelength to shift, which is why the excitation energy remains in the same position despite the shifted PL.

When a transverse magnetic field is applied in the Voigt direction, the samples exhibit characteristic oscillatory Faraday ellipticity signals, as shown in Figs.~\ref{fig:one}c-e for a field of 0.4\,T. These oscillations arise from the Larmor precession of electron and hole spins, with each carrier type contributing a distinct frequency component. The signals can be fitted and decomposed using:

\begin{equation}
S = \sum_i A_i \cos(\omega_i t) \exp\left(-\frac{t^2}{2T_{2,i}^{*2}}\right),
\end{equation}

\noindent where $i$ denotes the carrier type (electron or hole), $A_i$ is the amplitude, $\omega_i$ is the Larmor precession frequency, and $T_{2,i}^*$ is the spin dephasing time. The Larmor frequency is related to the applied magnetic field through:

\begin{equation}
\omega_i = \frac{g_i \mu_B B_V}{\hbar}
\end{equation}

\noindent where $g_i$ is the $g$-factor of the respective carrier, $\mu_B$ is the Bohr magneton, $B_V$ is the transverse magnetic field strength, and $\hbar$ is the reduced Planck constant. Holes have a higher effective mass than electrons, leading to more localized wavefunctions and correspondingly smaller $g$-factors, which results in lower precession frequencies.~\cite{Prechtel_2014, Sapienza_2016}.

A further result can be observed in Figure~\ref{fig:one}e, where, alongside the electron and hole precessions, there is an additional non-oscillating exponential decay element, which is also observed in the 8-layer sample investigated in Ref.~\cite{Evers_2024}. One possibility to explain it is related to the presence of a tilted magnetic field component, providing an additional Faraday geometry component, which causes the exponential decay. However, with further anisotropy measurements and the fact that this phenomenon is not observed in samples with 2 or 1 layers, we demonstrate that it is most probably related to inter-layer dot coupling and varying strain environment between the QD layers, which can affect heavy-hole and light-hole mixing.

\begin{figure}[]
\includegraphics[width=8cm]{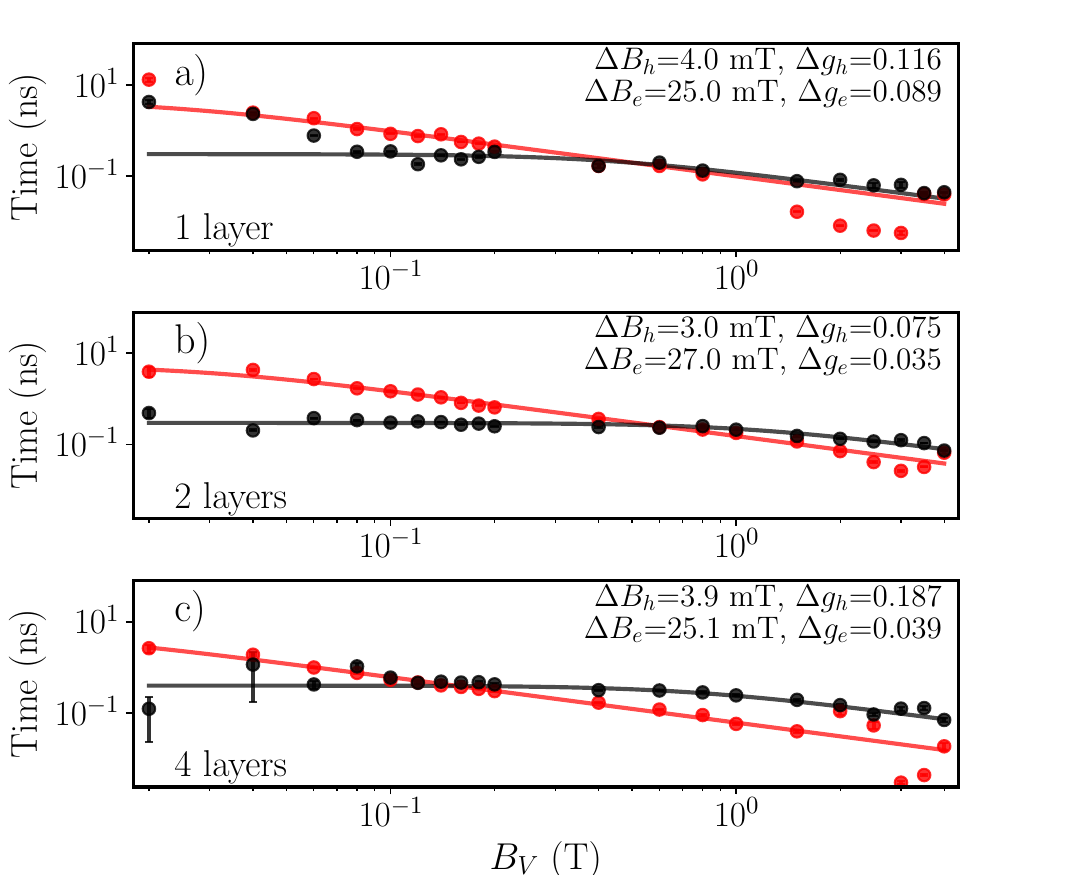}
\caption{Spin dephasing time ($T_2^*$) measurements as a function of transverse magnetic field ($B_V$) for InAs/InAlGaAs QD samples with different numbers of layers. The data is presented on log-log scales for samples containing 1, 2, and 4 QD layers (panels a, b, and c, respectively). Red circles represent hole spin dephasing times, and black circles represent electron spin dephasing times. The solid lines show fits to the experimental data using a model that accounts for $g$-factor spread and magnetic field fluctuations. The fitting parameters, including magnetic field spreads ($\Delta B_h$, $\Delta B_e$) and $g$-factor spreads ($\Delta g_h$, $\Delta g_e$) are displayed for each sample. The measurement configuration is the same as in Fig.~\ref{fig:one}.}
\label{fig:two}
\end{figure}

\begin{figure*}[]
\centering
\includegraphics[width=10cm]{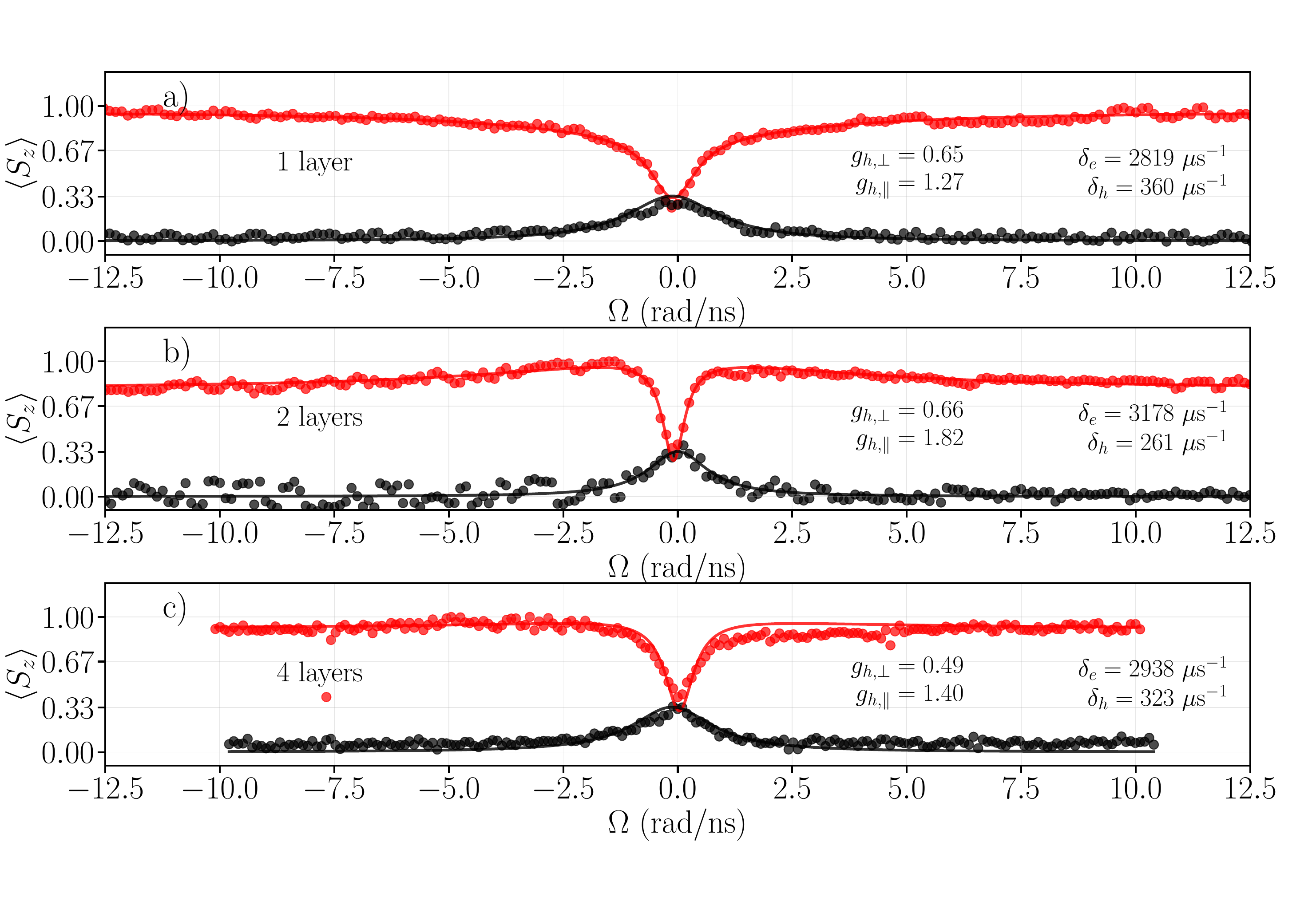}
\caption{Polarization recovery curves (red) and Hanle effect measurements (black) for (a) single-layer sample, (b) two-layer sample and (c) four-layer sample. The measurements show the total spin polarization $\langle S_z\rangle$ as a function of magnetic field strength. The fits for PRC are done with Eqs.~\ref{eq:P_function}. The pump power is 20~mW and probe power is 2~mW for both PRC and Hanle, for single and two-layer samples, and the pump power for four-layer sample is 10~mW. All measurements are performed at a temperature of 6\,K.}
\label{fig:three}
\end{figure*}

The spin dephasing times, displayed in Figures~\ref{fig:two}a-c, show a characteristic dependence on the applied magnetic field strength. At low magnetic fields, dephasing is dominated by hyperfine interactions with nuclear spins, whereas at higher fields, inhomogeneities in the QD ensemble, which cause a spread in the $g$-factor, become the limiting factor. This behavior is described by~\cite{Evers_2024}:

\begin{equation}
T_{2,i}^* = \frac{\hbar}{\sqrt{(\Delta g_i \mu_B B_V)^2 + (g_i \mu_B \Delta B_i)^2}}
\label{eq:dephasing}
\end{equation}

\noindent where $\Delta g_i$ represents the $g$-factor spread due to variations in QD size and composition, and $\Delta B_i$ characterizes the magnetic field fluctuations resulting from hyperfine interactions.


From the fits to the dephasing equation in Eq.~\ref{eq:dephasing}, we extract significantly different g-factor spreads for the three samples, shown in Table~\ref{tab:gfactor_spreads}. The single-layer sample shows the largest g-factor spreads for both electrons and holes, along with considerable magnetic field fluctuation spreads. The two-layer sample exhibits reduced g-factor spreads for both carrier types, with magnetic field fluctuation spreads remaining similar. The four-layer sample demonstrates a similar electron g-factor spread compared to the two-layer sample, but shows a significant increase in hole g-factor spread, while magnetic field fluctuation spreads remain comparable to the previous samples. The magnetic field fluctuation spreads, which reflect the strength of hyperfine interactions, show some variation between samples but remain in the same order of magnitude, suggesting similar nuclear spin environments in all three structures. There is more significant difference between our three samples and the eight-layer sample in Ref.~\cite{Evers_2024} in terms of magnetic field fluctuation values. The explanation for that is the measurement in Ref.~\cite{Evers_2024} extending only to 0.1\,T, inherently changing the fit results, since in our case we measure down to 0.02\,T.

\begin{figure*}[]
\centering
\includegraphics[width=10cm]{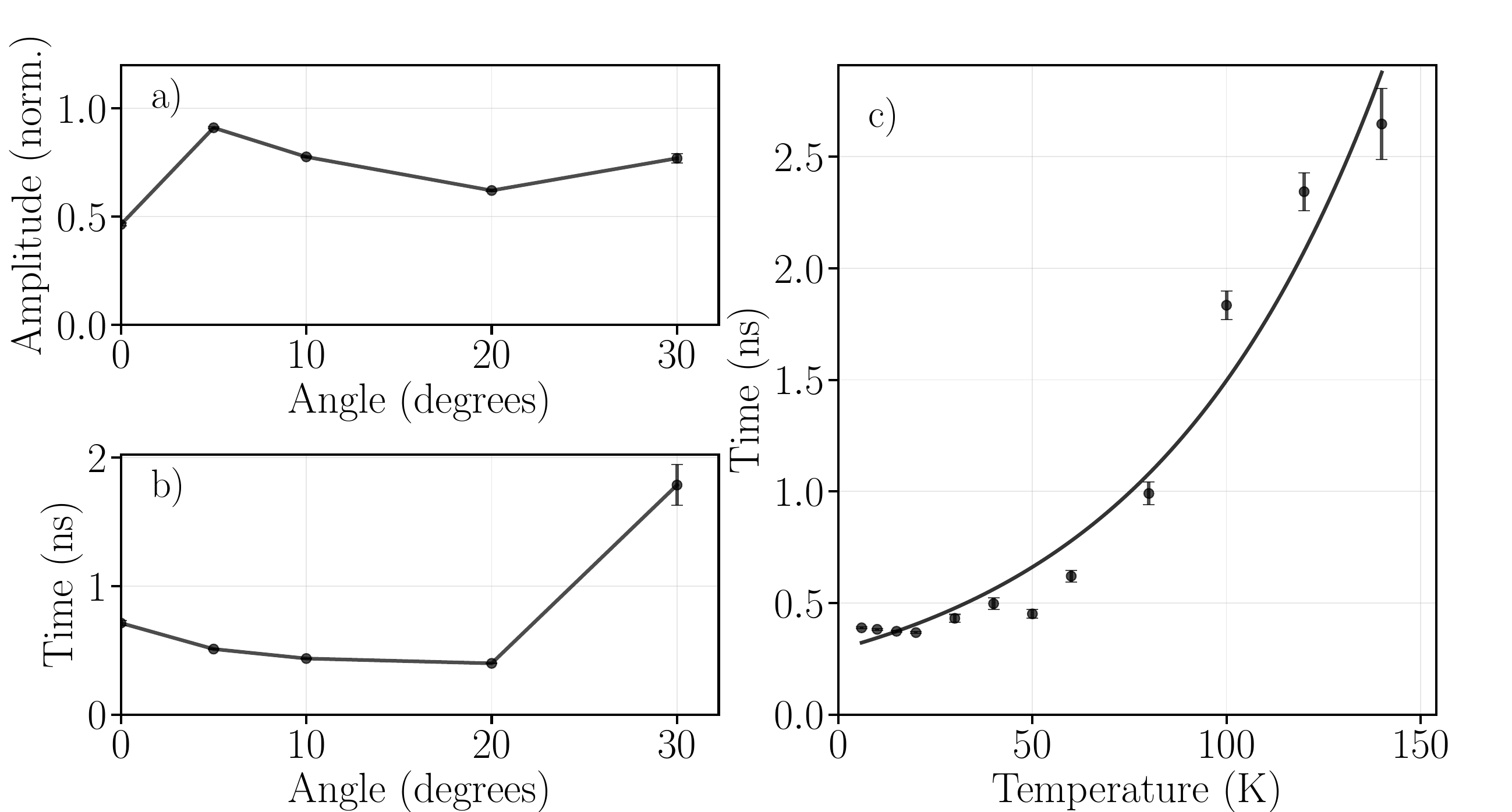}
\caption{Characterization of the additional exponential decay component observed in a four-layer QD sample. (a) Normalized amplitude of the exponential component as a function of magnetic field angle. (b) Decay time of the exponential component as a function of magnetic field angle. (c) Temperature dependence of the exponential component decay time. All measurements are performed at a temperature of 6~K (except for (c)), a transverse magnetic field strength of 0.4~T, a pump power of 10~mW, and a probe power of 2~mW.}
\label{fig:four}
\end{figure*}

\begin{table}[ht]
\centering
\begin{tabular}{|l|c|c|c|c|c|c|}
\hline
\textbf{Sample Type} & \textbf{$|g_e|$} & \textbf{$|g_h|$} & \textbf{$\Delta g_e$} & \textbf{$\Delta g_h$} & \textbf{$\Delta B_e$\,mT} & \textbf{$\Delta B_h$\,mT} \\
\hline
Single-layer & 1.51 & 0.64 & 0.09 & 0.12 & 25 & 4 \\
\hline
Two-layer & 1.42 & 0.72 & 0.04 & 0.08 & 27 & 3 \\
\hline
Four-layer & 1.15 & 0.49 & 0.04 & 0.19 & 25 & 4 \\
\hline
Eight-layer~\cite{Evers_2024} & 1.88 & 0.60 & 0.15 & 0.09 & 7 & 13 \\
\hline
\end{tabular}
\caption{Different sample g-factor values, g-factor spreads and magnetic field fluctuation spreads extracted from dephasing equation fits for electron and hole carriers across four QD layer structures.}
\label{tab:gfactor_spreads}
\end{table}

The $g$-factor spread shows more significant variation compared to the magnetic field spread. The reduced $g$-factor spread in the two-layer sample may indicate improved structural uniformity or different growth dynamics compared to the single-layer sample; however, the likely explanation is that the excitation energy remains the same while the PL for the two-layer sample blue shifts. This results in the effective excited PL width being significantly smaller compared to the single-layer sample, which means that the size distribution of the excited QDs is smaller than for the single-layer case, resulting in a smaller $g$-factor spread. Another contributing factor is that due to the PL blue shift, more higher energy QDs are excited. It has been demonstrated that higher energy transitions tend to have lower g-factor spreads~\cite{Belykh_2016}, which is in agreement with our reduced $g$-factor spread. A similar argument applies to the four-layer sample, where the PL is red-shifted and the excitation is at a higher energy than in the other two samples, increasing the hole $g$-factor spread. It is of interest to note that in both cases, the electron $g$-factor spread is smaller than that of the hole, and even decreases relative to the hole $g$-factor spread in the two-layer and four-layer structures. Compared to Ref.~\cite{Evers_2024}, it would be expected that the electron $g$-factor spread would become larger than for holes with an increasing layer amount. This is further addressed in the Discussion.

Lastly, the obtained transverse $g$-factors for the different samples can be seen in Table~\ref{tab:gfactor_spreads}. There is minimal change in the $g$-factors between one- and two-layer samples. However, the four-layer sample shows a significant decrease in $g$-factors. This is related to the different excitation energies between the samples. It is demonstrated that with increasing excitation energy, the electron $g$-factor decreases~\cite{Kim_2009, Belykh_2015}. In our case, the excitation energy is significantly increased in the four-layer sample, which results in a decreased electron $g$-factor. The same applies to the eight-layer sample in Ref.~\cite{Evers_2024}, where the $g$-factors are $|g_e| = 1.88$ and $|g_h| = 0.60$. The electron $g$-factor is significantly larger, which is expected, since the excitation done is at lower energies due to a red shift of the sample PL.



Figures~\ref{fig:three}a-c show polarization recovery curve (PRC) and Hanle measurements for the single-, two-, and four-layer samples, respectively. Both Hanle and PRC measurements are performed by measuring Faraday ellipticity immediately before pump and probe overlap, ensuring sufficient spin decay and measurements of the resident carriers. The main difference is that Hanle is performed in Voigt geometry, while PRC is conducted in Faraday geometry.

\begin{figure*}[]
\centering
\includegraphics[width=9cm]{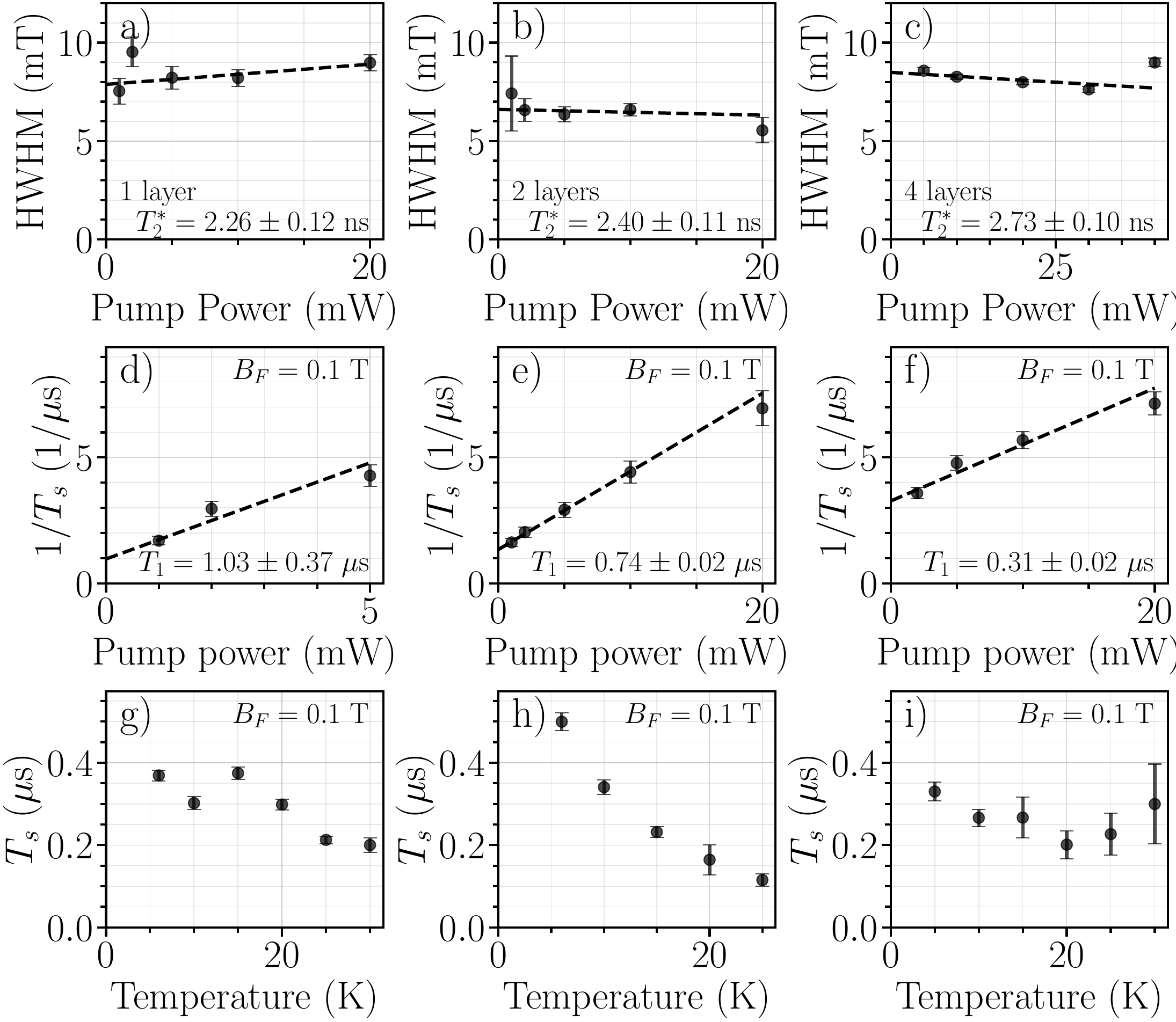}
\caption{Spin dynamics measurements for QD samples with different numbers of layers (1, 2, and 4 layers). Panels (a), (b), and (c) display the half-width at half maximum (HWHM) of Hanle curves as a function of pump power, from which spin dephasing times $T_2^*$ are extracted through linear extrapolation to zero pump power. Panels (d), (e), and (f) show the inverse spin lifetime (1/$T_s$) versus pump power measured under a longitudinal magnetic field of 0.1~T, with the intrinsic longitudinal spin relaxation times $T_1$ determined by extrapolating the linear fits to zero pump power. Panels (g), (h) and (i) show the spin lifetime in Faraday geometry dependency on the sample temperature, measured using the spin inertia method. All the samples have an applied magnetic field of 0.1~T. Pump power for the different panels is (g) 5~mW, (h) 2~mW, and (i) 2~mW. All measurements were performed at a temperature of 6\,K and a probe power of 2\,mW.}
\label{fig:five}
\end{figure*}

As mentioned in the Experimental Details, our sample is doped to provide electrons as resident carriers. It has been demonstrated previously that PRCs exhibit a V-like shape for n-doped samples and an M-shape for p-doped samples~\cite{Zhukov_2018}. From the single-layer sample PRC demonstrated in Fig.~\ref{fig:three}a, it can be seen that there are both electrons and holes as the resident carriers. However, in the two and four-layer samples (Fig.~\ref{fig:three}b), a more prominent M shape is observed, indicating that the electrons are observed as part of trions and holes are becoming resident carriers, showing that by adding additional layers, the number of holes as resident carriers increases. This can be attributed to electrons tunneling out to occupy lower-energy QDs, due to their lower effective mass~\cite{Mikhailov_2008, Evers_2024}, which leads to the optically induced hole doping.

The fits for both PRC and Hanle are done with Lorentzians that are provided by Ref.~\cite{Smirnov_2020}:
\begin{equation}
H(\Omega_L) \approx \frac{2}{3} \frac{\delta^2}{2\delta^2 + \Omega_L^2}
\label{eq:H_function}
\end{equation}

\begin{equation}
P(\Omega_L) \approx \frac{1}{3} \frac{2\delta^2 + 3\Omega_L^2}{2\delta^2 + \Omega_L^2},
\label{eq:P_function}
\end{equation}

\noindent where $P$ -- PRC, $H$ -- Hanle curve, $\delta$ -- dispersion of the nuclear fluctuations, and $\Omega_L = g \mu_B B/\hbar$. This model is applied by first fitting the Hanle curve with the known transverse $g$-factor to obtain the dispersion $\delta$, which is then used to fit the PRC's dispersion value and the carrier longitudinal $g$-factor. This basic model assumes that hyperfine interaction is isotropic and that the nuclear spin precession is significantly slower in comparison to electron spin precession. It should be noted that, due to the use of pulsed excitation, we are plotting total spin polarization rather than normalized spin polarization. In such case, the fit is $\langle S_z \rangle = S_0P(\Omega_L)$, where $S_0$ is the initial spin polarization~\cite{Smirnov_2020}. The reason for this is explained in Ref.~\cite{Smirnov_2020} Section V. In the single-layer sample, the PRC fitting yields electron dispersion values of $\delta_e = 2819 \,\mu s^{-1}$ and hole dispersion values of $\delta_h = 360 \ \mu s^{-1}$. For the two-layer sample, the electron dispersion slightly increases to $\delta_e = 3178 \ \mu s^{-1}$ and the obtained hole dispersion is $\delta_h = 261 \ \mu s^{-1}$. As for the four-layer sample, the dispersion values remain similar $\delta_e = 2938 \ \mu s^{-1}$ and $\delta_h = 323 \ \mu s^{-1}$. These widths indicate how strong the hyperfine interaction is~\cite{Smirnov_2020}, which means that for holes, the hyperfine interaction is significantly lower, which is expected since the nuclear interaction hyperfine constant of a hole is a magnitude smaller than for the electron~\cite{Testelin_2009}. Moreover, an offset is seen in the Hanle curve of the four-layer sample. This indicates that spin mode-locking is present, which we confirm and discuss in the text below. 

Returning to the aforementioned exponential decay component shown in Figure~\ref{fig:one}e, Figs.~\ref{fig:four}a and~\ref{fig:four}b prove that the additional component is not related to Faraday geometry. The sample was rotated, and a decaying, non-oscillating exponential component remained present at all magnetic field angles. It should be noted that, although the lowest decay time would be expected at a 0-degree angle due to the magnetic field being entirely in Voigt geometry, mechanical rotation of the sample introduces inaccuracies in determining the magnetic field angle.

Additionally, Fig.~\ref{fig:four}c shows the temperature dependence of the exponential decay time. In a magnetic field applied in the Faraday geometry, one would expect the measurement to reflect spin lifetime $T_1$, which typically decreases with increasing temperature~\cite{Marius_2025,Evers_2024,Mikhailov_2018}. Instead, the decay time in our case remains nearly constant up to 50\,K, after which it increases exponentially. A similar trend was reported in Ref.~\cite{Hostein_2008}, where exciton lifetimes were measured in QD molecules using time-resolved micro-photoluminescence. These observations support the conclusion that the exponential decay component originates when additional layers QD layers are added. However, this is not necessarily related to inter-dot coupling effects. This is further addressed in the Discussion.

Figure~\ref{fig:five} presents the different spin relaxation times for our samples. In Fig.~\ref{fig:five}a-c, Hanle measurements are performed at varying pump powers to extract the HWHM, from which the intrinsic spin dephasing time is determined by extrapolating to zero pump power using the following equation:

\begin{equation}
T_2^* = \frac{\hbar}{g\mu_B B_{\text{HWHM}}}.
\label{eq:T_s}
\end{equation}

\noindent The resulting spin dephasing times can be seen in Table~\ref{tab:spin_times}, showing only slight increase between the different samples. However, the obtained dephasing times are lower than the theoretical limit imposed by hyperfine interaction. To evaluate this limit, the hyperfine interaction strength can be calculated by fitting the PRC curves of all the samples with the following equation~\cite{Petrov_2008}:

\begin{equation}
    \bar{S}_z = S_0 \left[ \frac{1}{3} + \frac{2}{3} \frac{B_{\text{ext}}^2}{B_{\text{ext}}^2 + B_f^2} \right]
    \label{eq:nuclear_equal}
\end{equation}

\noindent where $B_\mathrm{ext}$ -- applied magnetic field, $S_0$ -- average electron-spin polarization at zero magnetic field, and $B_f$ -- averaged nuclear fluctuation field strength, which is half-width at half-minimum (HWHM) of the PRC dip. This equation is used to fit two components of the PRC, since for all samples we observe both a hole and an electron component. The results of the fits for the hole component of the single-layer sample are $B_f =5.2$~mT, $B_f=2.4$~mT for the two-layer sample, and $B_f=3.9$~mT for the four-layer sample. With this, the spin dephasing time limited by hyperfine interaction for holes can be calculated with the following equation~\cite{Greilich_2012}:

\begin{equation}
    T_2^* = 2\sqrt{3}\hbar / (g_h\mu_B B_f)
    \label{eq:dephase_hf}
\end{equation}

\noindent which results in 12\,ns for the single-layer sample, 23\,ns for the two-layer one and 20\,ns for the four-layer sample. The obtained $T_2^*$ values from Hanle measurements are significantly lower, indicating that effects beyond hyperfine interaction further reduce the dephasing times. One possibility is strain-induced effects that give rise to nuclear quadrupolar interactions, which can significantly reduce electron spin coherence times~\cite{Stockill_2016}. The inherent lattice mismatch in InAs/InAlGaAs QDs grown by SK creates structural asymmetries that enhance these quadrupolar effects. This interpretation is supported by recent studies on InAs/InGaAs/InP QDs grown by metalorganic vapour-phase epitaxy (MOVPE) droplet epitaxy, which demonstrate substantially improved spin dephasing times approaching the hyperfine limit due to the reduced strain achieved through this growth technique~\cite{Marius_2025}. It should be noted that these equations are usually applied for electrons rather than holes. Holes demonstrate significantly weaker hyperfine interactions compared to electrons~\cite{Testelin_2009}, so this is only an approximation; however, such spin dephasing times are expected for holes~\cite{Godden2012}.

Figures~\ref{fig:five}d-f show $T_1$ measurements obtained using the spin inertia method~\cite{Heisterkamp_2015}. In this measurement approach, a longitudinal magnetic field decouples carrier spins from nuclear spin fluctuations. With the probe delay fixed at a negative time, the pump modulation frequency is varied while monitoring the Faraday ellipticity signal. As the modulation period approaches the spin lifetime, the measured spin polarization diminishes, reducing the average signal amplitude according to the relation~\cite{Evers_2024}:

\begin{equation}
    S(f_m) = \frac{S_0}{\sqrt{1 + (2\pi f_m T_s)^2}},
    \label{eq:S}
\end{equation}

\noindent where $S_0$ -- initial amplitude of the Faraday ellipticity signal and $f_m$ -- pump beam polarization EOM modulation frequency. By measuring $T_s$ at different pump powers with a parallel magnetic field applied, a linear fit can be applied, and it can be extrapolated to obtain the intrinsic longitudinal spin lifetime $T_1$. The $T_1$ values across the samples can be seen in Table~\ref{tab:spin_times}. 

\begin{table}[ht]
\centering
\begin{tabular}{|l|c|c|}
\hline
\textbf{Sample Type} & \textbf{$T_2^*$ (ns)} & \textbf{$T_1$ ($\mu$s)} \\
\hline
Single-layer & 2.26 & 1.0 \\
\hline
Two-layer & 2.40 & 0.74 \\
\hline
Four-layer & 2.73 & 0.31\\
\hline
Eight-layer~\cite{Evers_2024} & - & 0.5 \\
\hline
\end{tabular}
\caption{Spin dephasing times ($T_2^*$) and longitudinal spin relaxation times ($T_1$) measured for electron carriers across the different QD layer structures. The $T_2^*$ value was not measured by Hanle in Ref.~\cite{Evers_2024}.}
\label{tab:spin_times}
\end{table}

The difference in $T_1$ times between single and two-layer samples, while significant, the error in the single-layer sample $T_1$ is large enough to put the value in the range of the two-layer sample. However, there is a further reduction for the four-layer sample compared to both the single and two-layer samples. The eight-layer sample in Ref.~\cite{Evers_2024} also exhibits a reduced $T_1$ of 0.5~$\mu$s, indicating a tendency for $T_1$ to decrease with additional layers. Figure~\ref{fig:three} demonstrates that with added layers, the resident carriers become holes; therefore, it is also safe to assume that the $T_1$ measurements are for holes when additional layers are added. This transition to hole-dominated charging contributes to the observed reduction in $T_1$ times, as holes exhibit enhanced spin-orbit coupling compared to electrons, which accelerates spin relaxation processes~\cite{Bulaev_2005, Chirolli_2008}. Moreover, the hole spin lifetime also shows a strong dependence on QD size. It was demonstrated that larger InAs/GaAs dots exhibit longer $T_1$ times~\cite{Wei_2010}. The particularly low $T_1$ observed in the four-layer sample can be understood through the interplay of these effects: the resident carriers are holes, which have an inherently shorter lifetime, and the sample was excited at a higher energy compared to the single and two-layer samples, which excite smaller QDs within the ensemble, leading to shorter $T_1$ times for holes. In contrast, the eight-layer sample in Ref.~\cite{Evers_2024} was excited at a lower energy, corresponding to larger QDs which have longer $T_1$ times, though this value remains reduced compared to single and two-layer samples due to the holes being the dominant carrier.

Figures~\ref{fig:five}g-i show the spin lifetime dependence on temperature with a magnetic field applied in Faraday geometry for single-, two-, and four-layer samples, respectively. The spin lifetime values in Fig.~\ref{fig:five}i are significantly lower because of using higher pump power, where it is known that increasing pump power speeds up spin relaxation due to photogenerated carriers, causing delocalization of resident carriers~\cite{Zhukov_2007}. Another factor is the intrinsically lower $T_1$ time for the four-layer structure, shown in Fig.~\ref{fig:five}f. The measurements were done up to 25-30~K due to the signal weakening as the temperature increases, making the fitting error too large at higher temperatures. The changing tendency between the different layer count structures is further addressed in the Discussion.


\begin{figure}[]
\includegraphics[width=8cm]{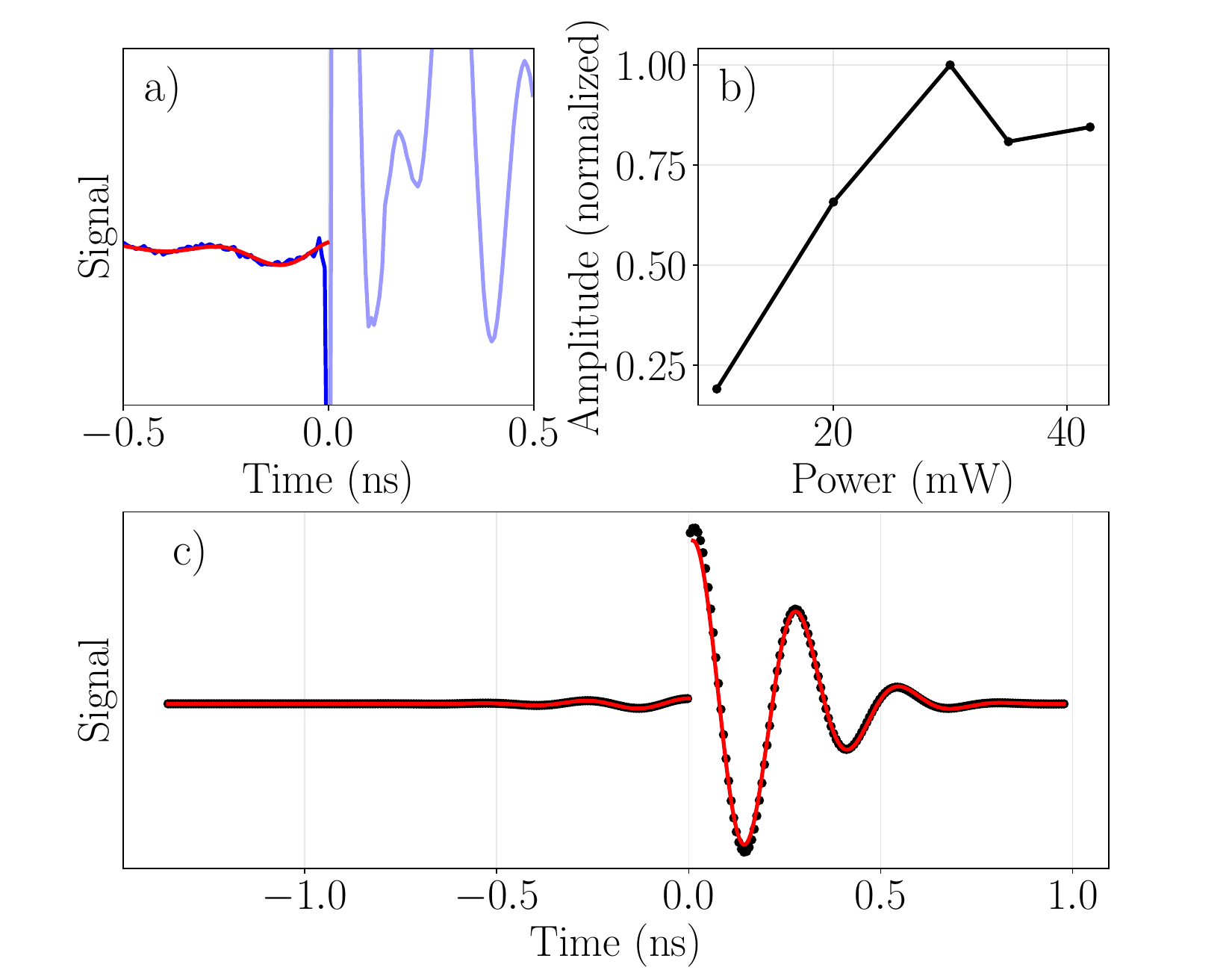}
\caption{(a) Time-resolved pump-probe signal for the four-layer sample, where the blue trace represents the raw data and the red line is a fit to the SML component. At negative time delays (left side), the signal shows the characteristic SML revival where spin coherence is recovered before the arrival of the pump pulse. At positive time delays (right side), normal spin precession is observed in Voigt geometry following optical excitation. (b) Normalized amplitude of the SML signal as a function of pump power, showing how the mode-locking effect strengthens with increasing excitation power until reaching saturation around 25-30\,mW. (c) The extracted SML data for holes (black data points) fitted with a theoretical model curve (red line). All the measurements were done at 6~K temperature, 0.4~T transverse magnetic field, and 2~mW probe power.}
\label{fig:six}
\end{figure}

Lastly, in Fig.~\ref{fig:six}a, the zoomed-in signal of the four-layer sample at negative time delay is displayed. In this case, a weak, single-frequency oscillation can be observed. This effect is called spin mode-locking (SML). In SML, the coherent spin dynamics are sustained across multiple pump pulse cycles through a feedback mechanism. When the spin coherence time $T_2$ exceeds the laser repetition period $T_R$, spins that precess commensurate with the pulse repetition frequency can maintain phase coherence between successive excitation pulses~\cite{Evers_2024, Greilich_2006}. This manifests as a revival of the spin signal at negative time delays, where the probe pulse arrives before the pump pulse. In this case, the SML signal frequency corresponds to holes, since the SML signal frequency is the same as the hole precession frequency seen at positive time delays.

The existence of SML allows us to estimate $T_2$ for the four-layer sample. The hole spin coherence time $T_2$ was extracted by modeling the SML signal amplitude ratio before and after pump pulse excitation. The modeling incorporated experimentally determined hole $g$-factors, $g$-factor spreads ($\Delta g_h$), and assumed a Gaussian distribution of $g$-factors across the QD ensemble. Refs.~\cite{Yugova_2009, Yugova_2012, Evers_2024, Greilich_2006} go into more details about how the modeling is done.

In Ref.~\cite{Evers_2024}, the authors were only able to estimate the range of $T_2$ times due to not observing SML amplitude saturation or Rabi oscillations. In our case, we can determine it much more accurately due to observing saturation of the SML amplitude, as can be seen in Fig.~\ref{fig:six}b. To determine it, we model the QD ensemble using a Gaussian $g$-factor distribution with a width of $\Delta g_h=0.11$ and $g$-factor $|g_h|=0.65$. The amplitude ratio of SML and the amplitude of the signal after excitation then depends on $T_2$ time and the pulse area $\Theta$. Since we observe saturation, the pulse area is $\Theta=\pi$, and thus the coherence time is $T_2=13~\text{ns}$. The plotted model can be seen in Fig.~\ref{fig:six}c.



\section{Discussion}

Several noteworthy effects emerge upon incorporating additional QD layers with narrow barrier separations. First, the n-doped sample becomes p-doped when a second layer is added, and with four layers, there is indication that there is again more electron carriers, which is possible due to variances in doping between samples, but holes are still present, as shown in Figs.~\ref{fig:three}a-c. In the previous section, we mentioned that the cause of this is the tunneling of excited electrons to lower-energy QDs. The reason for this is likely that upon growing additional layers, the QDs on that layer have lower energy. This has been demonstrated for InAs/GaAs QDs, where as layers are stacked, the QD size increases since the strain distribution from the underlying layers causes the preferential localization of QDs on top~\cite{Krenner_2005, Usman_2011}. Moreover, while the barrier thickness is 15\,nm, the barrier is not on top of the QDs, but instead covers them. The QD height is also between 9-13~nm~\cite{Belykh_2016}, which makes the effective distance between the QDs short enough for tunneling to occur. 

In addition to tunneling, we see an additional non-oscillating decaying exponential starting to appear in the four-layer structure, seen in Fig.~\ref{fig:one}d, which is also seen in the eight-layer structure in Ref.~\cite{Evers_2024} Fig.~2a. In the referenced paper, they attribute this additional component to either a molecular or excitonic state caused by coupling between the QDs. In our case, we also attribute it to coupling-related effects, since we exclude the possibility of the exponential resulting from Faraday geometry by performing anisotropy measurements shown in Figs.~\ref{fig:four}a-b, and also due to the fact that we do not see the additional component in either single or two-layer samples.

We hypothesize that there are two effects associated with this non-oscillating component. It is known that, due to QD confinement, there is significant light-hole (LH) and heavy-hole (HH) mixing~\cite{Luo_2014}, which affects the optical and spin properties of QDs, in particular, causing significant $g$-factor anisotropy and non-zero in-plane $g$-factor values for the corresponding hole component\cite{Belykh_2016_anisotropy}. However, for some quantum well samples, a pure HH state has been observed, which shows transverse $g$-factor values near zero~\cite{Marie_1999} or even effectively zero when considering thin quantum wells, in which case the splitting between the LH and HH is relatively large when ignoring cubic symmetry~\cite{Semina_2023}. Therefore, we attribute the non-oscillating component to potentially reduced heavy-hole and light-hole mixing, which is caused by the addition of multiple layers that change the strain environment and, in turn, reduce symmetry, resulting in more significant splitting between LH and HH. This is further confirmed by examining the eight-layer structure in Ref.~\cite{Evers_2024}, where the non-oscillating component exhibits an increase in both decay time and amplitude, suggesting that additional layers may further reduce HH and LH mixing.

Subsequently, Fig.~\ref{fig:four}c demonstrates an unexpected result where both the electron and the potential HH dephasing times increase with temperature. These results show strong similarity to time-resolved microphotoluminescence exciton lifetime measurements in InAs/InP QDs~\cite{Hostein_2008} and in InAs/GaAs QDs~\cite{Yu_1996,Fiore_2000,Wang_1994}, where the lifetime increases with temperature due to the carriers being excited to p-shell states with increased temperature, causing lowered radiative rates since the carriers have to relax back to s-shell before recombining. Moreover, indirect excitons in coupled QD systems have been demonstrated with photoluminescence measurements, by using an electric field to put the electron and hole in resonance~\cite{Daniels_2013}. Therefore, as we have confirmed that tunneling in our sample is a possibility due to the effective distances between the QDs and due to the results shown in Fig.~\ref{fig:three}, we then assume that the increasing dephasing times of exciton and heavy-hole seen in Fig.~\ref{fig:four}c is due to the formation of an indirect exciton.

In our case, we do not need to do electric field tuning due to our measurement observing spin polarization, rather than photoluminescence. However, the previously discussed measurements in Refs.~\cite{Hostein_2008, Yu_1996, Fiore_2000, Wang_1994} show increased exciton lifetime, rather than investigating spin dephasing. The observed dephasing pattern arises because holes in indirect excitons experience primarily hyperfine interactions as their dominant dephasing mechanism~\cite{Rautert_2019}. This hyperfine interaction-limited dephasing correlates the hole spin dephasing time directly to the exciton lifetime. Therefore, when temperature increases, exciton lifetimes extend, and the dephasing time increases proportionally. As to why this is not observed for the direct excitons in either the four-layer or lower count layer samples, this is due to the indirect excitons having an inherently longer lifetime because of their reduced wavefunction overlap~\cite{Sivalertporn_2011}, making them more affected by the aforementioned radiative overlap.

Another emerging effect due to multiple layers of QDs can be seen in Figs.~\ref{fig:five}g-i. In the Figs.~\ref{fig:five}g and h, the tendencies are expected, where in panel g, a more or less constant spin lifetime is observed up to 20~K and then starts reducing, which indicates that the resident carriers are electrons~\cite{Mikhailov_2018}. In plot h, the lifetime starts reducing immediately at 10~K, which suggests that the resident carriers are then holes~\cite{Mikhailov_2018}. The PRC results in Figs.~\ref{fig:three}a and b further confirm these two plots. In the case of Fig.~\ref{fig:five}i, it is expected that it would be the same as in plot h, due to the resident carriers being holes; however, the spin lifetime remains nearly constant through the whole range. Almost identical behavior is seen in the eight-layer sample investigated in Ref.~\cite{Evers_2024}, showing that this effect, just like SML and the non-oscillating component, only appears at 4 layers or more. 

Lastly, along with the additional component appearing with the four-layer sample, we also see spin mode-locking. Moreover, we can see improvements in $T_2$ times going from four-layer sample to eight-layer sample shown in Ref.~\cite{Evers_2024}, however, they are quite marginal and in this case are attributed to error. As to the reason of SML appearance, it can be explained by the aforementioned tunneling. As more and more layers are added, there are more QDs for electrons to tunnel to, leaving more holes behind. This greatly increases the hole population, which allows us to observe SML. As for the $T_2$ values themselves, they are relatively low compared to $T_2=1.1\,\mu s$ of AlAs/GaAs QDs~\cite{Greve_2011}. However, hole spin coherence times have not been previously observed in telecom QDs. Therefore, this shows that multi-layer QD structures have potential for quantim information applications. To get a better insight not only in to $T_2$ but also other phenomena we have observed in this paper, additional measurements and analysis is needed.






\flushbottom

\section{Conclusions}
This layer-resolved study identifies how vertical stacking of MBE grown telecom C-band QDs impacts spin properties, revealing several phenomena that emerge with increasing layer count in InAs/InAlGaAs QD configurations. The transition from electron-dominated to hole-dominated resident carriers with increasing layers represents a fundamental shift in spin dynamics, where single-layer samples exhibit both electron and hole components while two- and four-layer samples show predominantly hole-like behavior due to optically induced electron tunneling to lower-energy dots in additional layers. Four-layer structures exhibit distinctive inter-layer coupling effects including an additional non-oscillating exponential decay component that persists across all magnetic field angles with an unusual temperature dependence where spin relaxation times increase rather than decrease with temperature, attributed to indirect exciton effects. Also, the emergence of spin mode locking that enables direct measurement of hole coherence times $T_2 \approx 13$\,ns. While transverse dephasing times $T_2^*$ remain relatively stable across layer counts (2.26-2.73\,ns), they fall short of hyperfine-limited theoretical values, suggesting additional dephasing mechanisms such as strain-induced quadrupolar interactions, and the reduction in longitudinal relaxation times $T_1$ with increasing layers (1.03\,$\mu$s down to 0.31\,$\mu$s) highlights the trade-off between optical signal enhancement and spin lifetime preservation. These measurements establish the layer-dependent evolution of spin properties in vertically coupled telecom QDs, demonstrating that interlayer coupling introduces both new measurement capabilities and constraints on spin coherence parameters.

\section*{Acknowledgments}
We thank N.~E.~Kopteva for valuable discussions and help with the SML model. Data is made available upon request.

\bibliography{references} 

\begin{thebibliography}{47}%
\makeatletter
\providecommand \@ifxundefined [1]{%
 \@ifx{#1\undefined}
}%
\providecommand \@ifnum [1]{%
 \ifnum #1\expandafter \@firstoftwo
 \else \expandafter \@secondoftwo
 \fi
}%
\providecommand \@ifx [1]{%
 \ifx #1\expandafter \@firstoftwo
 \else \expandafter \@secondoftwo
 \fi
}%
\providecommand \natexlab [1]{#1}%
\providecommand \enquote  [1]{``#1''}%
\providecommand \bibnamefont  [1]{#1}%
\providecommand \bibfnamefont [1]{#1}%
\providecommand \citenamefont [1]{#1}%
\providecommand \href@noop [0]{\@secondoftwo}%
\providecommand \href [0]{\begingroup \@sanitize@url \@href}%
\providecommand \@href[1]{\@@startlink{#1}\@@href}%
\providecommand \@@href[1]{\endgroup#1\@@endlink}%
\providecommand \@sanitize@url [0]{\catcode `\\12\catcode `\$12\catcode `\&12\catcode `\#12\catcode `\^12\catcode `\_12\catcode `\%12\relax}%
\providecommand \@@startlink[1]{}%
\providecommand \@@endlink[0]{}%
\providecommand \url  [0]{\begingroup\@sanitize@url \@url }%
\providecommand \@url [1]{\endgroup\@href {#1}{\urlprefix }}%
\providecommand \urlprefix  [0]{URL }%
\providecommand \Eprint [0]{\href }%
\providecommand \doibase [0]{https://doi.org/}%
\providecommand \selectlanguage [0]{\@gobble}%
\providecommand \bibinfo  [0]{\@secondoftwo}%
\providecommand \bibfield  [0]{\@secondoftwo}%
\providecommand \translation [1]{[#1]}%
\providecommand \BibitemOpen [0]{}%
\providecommand \bibitemStop [0]{}%
\providecommand \bibitemNoStop [0]{.\EOS\space}%
\providecommand \EOS [0]{\spacefactor3000\relax}%
\providecommand \BibitemShut  [1]{\csname bibitem#1\endcsname}%
\let\auto@bib@innerbib\@empty
\bibitem [{\citenamefont {Vajner}\ \emph {et~al.}(2022)\citenamefont {Vajner}, \citenamefont {Rickert}, \citenamefont {Gao}, \citenamefont {Kaymazlar},\ and\ \citenamefont {Heindel}}]{Vajner_2022}%
  \BibitemOpen
  \bibfield  {author} {\bibinfo {author} {\bibfnamefont {D.~A.}\ \bibnamefont {Vajner}}, \bibinfo {author} {\bibfnamefont {L.}~\bibnamefont {Rickert}}, \bibinfo {author} {\bibfnamefont {T.}~\bibnamefont {Gao}}, \bibinfo {author} {\bibfnamefont {K.}~\bibnamefont {Kaymazlar}},\ and\ \bibinfo {author} {\bibfnamefont {T.}~\bibnamefont {Heindel}},\ }\bibfield  {title} {\bibinfo {title} {Quantum communication using semiconductor quantum dots},\ }\href {https://doi.org/https://doi.org/10.1002/qute.202100116} {\bibfield  {journal} {\bibinfo  {journal} {Advanced Quantum Technologies}\ }\textbf {\bibinfo {volume} {5}},\ \bibinfo {pages} {2100116} (\bibinfo {year} {2022})}\BibitemShut {NoStop}%
\bibitem [{\citenamefont {Miyazawa}\ \emph {et~al.}(2016)\citenamefont {Miyazawa}, \citenamefont {Takemoto}, \citenamefont {Nambu}, \citenamefont {Miki}, \citenamefont {Yamashita}, \citenamefont {Terai}, \citenamefont {Fujiwara}, \citenamefont {Sasaki}, \citenamefont {Sakuma}, \citenamefont {Takatsu}, \citenamefont {Yamamoto},\ and\ \citenamefont {Arakawa}}]{Miyazawa_2016}%
  \BibitemOpen
  \bibfield  {author} {\bibinfo {author} {\bibfnamefont {T.}~\bibnamefont {Miyazawa}}, \bibinfo {author} {\bibfnamefont {K.}~\bibnamefont {Takemoto}}, \bibinfo {author} {\bibfnamefont {Y.}~\bibnamefont {Nambu}}, \bibinfo {author} {\bibfnamefont {S.}~\bibnamefont {Miki}}, \bibinfo {author} {\bibfnamefont {T.}~\bibnamefont {Yamashita}}, \bibinfo {author} {\bibfnamefont {H.}~\bibnamefont {Terai}}, \bibinfo {author} {\bibfnamefont {M.}~\bibnamefont {Fujiwara}}, \bibinfo {author} {\bibfnamefont {M.}~\bibnamefont {Sasaki}}, \bibinfo {author} {\bibfnamefont {Y.}~\bibnamefont {Sakuma}}, \bibinfo {author} {\bibfnamefont {M.}~\bibnamefont {Takatsu}}, \bibinfo {author} {\bibfnamefont {T.}~\bibnamefont {Yamamoto}},\ and\ \bibinfo {author} {\bibfnamefont {Y.}~\bibnamefont {Arakawa}},\ }\bibfield  {title} {\bibinfo {title} {Single-photon emission at 1.5 \si{\micro\meter} from an {InAs/InP} quantum dot with highly suppressed multi-photon emission probabilities},\ }\href {https://doi.org/10.1063/1.4961888} {\bibfield
  {journal} {\bibinfo  {journal} {Applied Physics Letters}\ }\textbf {\bibinfo {volume} {109}},\ \bibinfo {pages} {132106} (\bibinfo {year} {2016})}\BibitemShut {NoStop}%
\bibitem [{\citenamefont {Holewa}\ \emph {et~al.}(2021)\citenamefont {Holewa}, \citenamefont {Sakanas}, \citenamefont {Gur}, \citenamefont {Mrowi'nski}, \citenamefont {Huck}, \citenamefont {Wang}, \citenamefont {Musiał}, \citenamefont {Yvind}, \citenamefont {Gregersen}, \citenamefont {Syperek},\ and\ \citenamefont {Semenova}}]{Pawel_2021}%
  \BibitemOpen
  \bibfield  {author} {\bibinfo {author} {\bibfnamefont {P.}~\bibnamefont {Holewa}}, \bibinfo {author} {\bibfnamefont {A.}~\bibnamefont {Sakanas}}, \bibinfo {author} {\bibfnamefont {U.~M.}\ \bibnamefont {Gur}}, \bibinfo {author} {\bibfnamefont {P.}~\bibnamefont {Mrowi'nski}}, \bibinfo {author} {\bibfnamefont {A.}~\bibnamefont {Huck}}, \bibinfo {author} {\bibfnamefont {B.-Y.}\ \bibnamefont {Wang}}, \bibinfo {author} {\bibfnamefont {A.}~\bibnamefont {Musiał}}, \bibinfo {author} {\bibfnamefont {K.}~\bibnamefont {Yvind}}, \bibinfo {author} {\bibfnamefont {N.}~\bibnamefont {Gregersen}}, \bibinfo {author} {\bibfnamefont {M.}~\bibnamefont {Syperek}},\ and\ \bibinfo {author} {\bibfnamefont {E.}~\bibnamefont {Semenova}},\ }\bibfield  {title} {\bibinfo {title} {Bright quantum dot single-photon emitters at telecom bands heterogeneously integrated on {Si}},\ }\href {https://api.semanticscholar.org/CorpusID:233240645} {\bibfield  {journal} {\bibinfo  {journal} {ACS Photonics}\ }\textbf {\bibinfo {volume} {9}},\ \bibinfo
  {pages} {2273 } (\bibinfo {year} {2021})}\BibitemShut {NoStop}%
\bibitem [{\citenamefont {Bauer}\ \emph {et~al.}(2021)\citenamefont {Bauer}, \citenamefont {Wang}, \citenamefont {Hoppe}, \citenamefont {Nawrath}, \citenamefont {Fischer}, \citenamefont {Witz}, \citenamefont {Kaschel}, \citenamefont {Schweikert}, \citenamefont {Jetter}, \citenamefont {Portalupi}, \citenamefont {Berroth},\ and\ \citenamefont {Michler}}]{Bauer_2021}%
  \BibitemOpen
  \bibfield  {author} {\bibinfo {author} {\bibfnamefont {S.}~\bibnamefont {Bauer}}, \bibinfo {author} {\bibfnamefont {D.}~\bibnamefont {Wang}}, \bibinfo {author} {\bibfnamefont {N.}~\bibnamefont {Hoppe}}, \bibinfo {author} {\bibfnamefont {C.}~\bibnamefont {Nawrath}}, \bibinfo {author} {\bibfnamefont {J.}~\bibnamefont {Fischer}}, \bibinfo {author} {\bibfnamefont {N.}~\bibnamefont {Witz}}, \bibinfo {author} {\bibfnamefont {M.}~\bibnamefont {Kaschel}}, \bibinfo {author} {\bibfnamefont {C.}~\bibnamefont {Schweikert}}, \bibinfo {author} {\bibfnamefont {M.}~\bibnamefont {Jetter}}, \bibinfo {author} {\bibfnamefont {S.~L.}\ \bibnamefont {Portalupi}}, \bibinfo {author} {\bibfnamefont {M.}~\bibnamefont {Berroth}},\ and\ \bibinfo {author} {\bibfnamefont {P.}~\bibnamefont {Michler}},\ }\bibfield  {title} {\bibinfo {title} {Achieving stable fiber coupling of quantum dot telecom c-band single-photons to an {SOI} photonic device},\ }\href {https://doi.org/10.1063/5.0067749} {\bibfield  {journal} {\bibinfo  {journal}
  {Applied Physics Letters}\ }\textbf {\bibinfo {volume} {119}},\ \bibinfo {pages} {211101} (\bibinfo {year} {2021})}\BibitemShut {NoStop}%
\bibitem [{\citenamefont {Rahaman}\ \emph {et~al.}(2024)\citenamefont {Rahaman}, \citenamefont {Harper}, \citenamefont {Lee}, \citenamefont {Kim}, \citenamefont {Buyukkaya}, \citenamefont {Patel}, \citenamefont {Hawkins}, \citenamefont {Kim}, \citenamefont {Addamane},\ and\ \citenamefont {Waks}}]{Rahaman_2024}%
  \BibitemOpen
  \bibfield  {author} {\bibinfo {author} {\bibfnamefont {M.~H.}\ \bibnamefont {Rahaman}}, \bibinfo {author} {\bibfnamefont {S.}~\bibnamefont {Harper}}, \bibinfo {author} {\bibfnamefont {C.-M.}\ \bibnamefont {Lee}}, \bibinfo {author} {\bibfnamefont {K.-Y.}\ \bibnamefont {Kim}}, \bibinfo {author} {\bibfnamefont {M.}~\bibnamefont {Buyukkaya}}, \bibinfo {author} {\bibfnamefont {V.}~\bibnamefont {Patel}}, \bibinfo {author} {\bibfnamefont {S.}~\bibnamefont {Hawkins}}, \bibinfo {author} {\bibfnamefont {J.-H.}\ \bibnamefont {Kim}}, \bibinfo {author} {\bibfnamefont {S.}~\bibnamefont {Addamane}},\ and\ \bibinfo {author} {\bibfnamefont {E.}~\bibnamefont {Waks}},\ }\bibfield  {title} {\bibinfo {title} {Efficient, indistinguishable telecom c-band photons using a tapered nanobeam waveguide},\ }\href {https://doi.org/10.1021/acsphotonics.4c00625} {\bibfield  {journal} {\bibinfo  {journal} {ACS Photonics}\ }\textbf {\bibinfo {volume} {11}},\ \bibinfo {pages} {2738–2744} (\bibinfo {year} {2024})}\BibitemShut {NoStop}%
\bibitem [{\citenamefont {Mikhailov}\ \emph {et~al.}(2018)\citenamefont {Mikhailov}, \citenamefont {Belykh}, \citenamefont {Yakovlev}, \citenamefont {Grigoryev}, \citenamefont {Reithmaier}, \citenamefont {Benyoucef},\ and\ \citenamefont {Bayer}}]{Mikhailov_2018}%
  \BibitemOpen
  \bibfield  {author} {\bibinfo {author} {\bibfnamefont {A.~V.}\ \bibnamefont {Mikhailov}}, \bibinfo {author} {\bibfnamefont {V.~V.}\ \bibnamefont {Belykh}}, \bibinfo {author} {\bibfnamefont {D.~R.}\ \bibnamefont {Yakovlev}}, \bibinfo {author} {\bibfnamefont {P.~S.}\ \bibnamefont {Grigoryev}}, \bibinfo {author} {\bibfnamefont {J.~P.}\ \bibnamefont {Reithmaier}}, \bibinfo {author} {\bibfnamefont {M.}~\bibnamefont {Benyoucef}},\ and\ \bibinfo {author} {\bibfnamefont {M.}~\bibnamefont {Bayer}},\ }\bibfield  {title} {\bibinfo {title} {Electron and hole spin relaxation in {InP}-based self-assembled quantum dots emitting at telecom wavelengths},\ }\href {https://doi.org/10.1103/PhysRevB.98.205306} {\bibfield  {journal} {\bibinfo  {journal} {Phys. Rev. B}\ }\textbf {\bibinfo {volume} {98}},\ \bibinfo {pages} {205306} (\bibinfo {year} {2018})}\BibitemShut {NoStop}%
\bibitem [{\citenamefont {Evers}\ \emph {et~al.}(2025)\citenamefont {Evers}, \citenamefont {Kopteva}, \citenamefont {Nedelea}, \citenamefont {Kors}, \citenamefont {Kaur}, \citenamefont {Peter~Reithmaier}, \citenamefont {Benyoucef}, \citenamefont {Bayer},\ and\ \citenamefont {Greilich}}]{Evers_2024}%
  \BibitemOpen
  \bibfield  {author} {\bibinfo {author} {\bibfnamefont {E.}~\bibnamefont {Evers}}, \bibinfo {author} {\bibfnamefont {N.~E.}\ \bibnamefont {Kopteva}}, \bibinfo {author} {\bibfnamefont {V.}~\bibnamefont {Nedelea}}, \bibinfo {author} {\bibfnamefont {A.}~\bibnamefont {Kors}}, \bibinfo {author} {\bibfnamefont {R.}~\bibnamefont {Kaur}}, \bibinfo {author} {\bibfnamefont {J.}~\bibnamefont {Peter~Reithmaier}}, \bibinfo {author} {\bibfnamefont {M.}~\bibnamefont {Benyoucef}}, \bibinfo {author} {\bibfnamefont {M.}~\bibnamefont {Bayer}},\ and\ \bibinfo {author} {\bibfnamefont {A.}~\bibnamefont {Greilich}},\ }\bibfield  {title} {\bibinfo {title} {Hole spin coherence in {InAs/InAlGaAs} self-assembled quantum dots emitting at telecom wavelengths},\ }\href {https://doi.org/https://doi.org/10.1002/pssb.202400174} {\bibfield  {journal} {\bibinfo  {journal} {physica status solidi (b)}\ }\textbf {\bibinfo {volume} {262}},\ \bibinfo {pages} {2400174} (\bibinfo {year} {2025})}\BibitemShut {NoStop}%
\bibitem [{\citenamefont {Smirnov}\ \emph {et~al.}(2020)\citenamefont {Smirnov}, \citenamefont {Zhukov}, \citenamefont {Yakovlev}, \citenamefont {Kirstein}, \citenamefont {Bayer},\ and\ \citenamefont {Greilich}}]{Smirnov_2020}%
  \BibitemOpen
  \bibfield  {author} {\bibinfo {author} {\bibfnamefont {D.~S.}\ \bibnamefont {Smirnov}}, \bibinfo {author} {\bibfnamefont {E.~A.}\ \bibnamefont {Zhukov}}, \bibinfo {author} {\bibfnamefont {D.~R.}\ \bibnamefont {Yakovlev}}, \bibinfo {author} {\bibfnamefont {E.}~\bibnamefont {Kirstein}}, \bibinfo {author} {\bibfnamefont {M.}~\bibnamefont {Bayer}},\ and\ \bibinfo {author} {\bibfnamefont {A.}~\bibnamefont {Greilich}},\ }\bibfield  {title} {\bibinfo {title} {Spin polarization recovery and hanle effect for charge carriers interacting with nuclear spins in semiconductors},\ }\href {https://doi.org/10.1103/PhysRevB.102.235413} {\bibfield  {journal} {\bibinfo  {journal} {Phys. Rev. B}\ }\textbf {\bibinfo {volume} {102}},\ \bibinfo {pages} {235413} (\bibinfo {year} {2020})}\BibitemShut {NoStop}%
\bibitem [{\citenamefont {Li}\ \emph {et~al.}(2023)\citenamefont {Li}, \citenamefont {Xu},\ and\ \citenamefont {Zhang}}]{Li_2023}%
  \BibitemOpen
  \bibfield  {author} {\bibinfo {author} {\bibfnamefont {X.}~\bibnamefont {Li}}, \bibinfo {author} {\bibfnamefont {Q.}~\bibnamefont {Xu}},\ and\ \bibinfo {author} {\bibfnamefont {Z.}~\bibnamefont {Zhang}},\ }\bibfield  {title} {\bibinfo {title} {Molecular beam epitaxy growth of quantum wires and quantum dots},\ }\href {https://doi.org/10.3390/nano13060960} {\bibfield  {journal} {\bibinfo  {journal} {Nanomaterials}\ }\textbf {\bibinfo {volume} {13}},\ \bibinfo {pages} {960} (\bibinfo {year} {2023})}\BibitemShut {NoStop}%
\bibitem [{\citenamefont {Stinaff}\ \emph {et~al.}(2006)\citenamefont {Stinaff}, \citenamefont {Scheibner}, \citenamefont {Bracker}, \citenamefont {Ponomarev}, \citenamefont {Korenev}, \citenamefont {Ware}, \citenamefont {Doty}, \citenamefont {Reinecke},\ and\ \citenamefont {Gammon}}]{Stinaff2006}%
  \BibitemOpen
  \bibfield  {author} {\bibinfo {author} {\bibfnamefont {E.~A.}\ \bibnamefont {Stinaff}}, \bibinfo {author} {\bibfnamefont {M.}~\bibnamefont {Scheibner}}, \bibinfo {author} {\bibfnamefont {A.~S.}\ \bibnamefont {Bracker}}, \bibinfo {author} {\bibfnamefont {I.~V.}\ \bibnamefont {Ponomarev}}, \bibinfo {author} {\bibfnamefont {V.~L.}\ \bibnamefont {Korenev}}, \bibinfo {author} {\bibfnamefont {M.~E.}\ \bibnamefont {Ware}}, \bibinfo {author} {\bibfnamefont {M.~F.}\ \bibnamefont {Doty}}, \bibinfo {author} {\bibfnamefont {T.~L.}\ \bibnamefont {Reinecke}},\ and\ \bibinfo {author} {\bibfnamefont {D.}~\bibnamefont {Gammon}},\ }\bibfield  {title} {\bibinfo {title} {Optical signatures of coupled quantum dots},\ }\href {https://doi.org/10.1126/science.1121184} {\bibfield  {journal} {\bibinfo  {journal} {Science}\ }\textbf {\bibinfo {volume} {311}},\ \bibinfo {pages} {636} (\bibinfo {year} {2006})}\BibitemShut {NoStop}%
\bibitem [{\citenamefont {Kim}\ \emph {et~al.}(2008)\citenamefont {Kim}, \citenamefont {Economou}, \citenamefont {B\u{a}descu}, \citenamefont {Scheibner}, \citenamefont {Bracker}, \citenamefont {Bashkansky}, \citenamefont {Reinecke},\ and\ \citenamefont {Gammon}}]{Kim2008}%
  \BibitemOpen
  \bibfield  {author} {\bibinfo {author} {\bibfnamefont {D.}~\bibnamefont {Kim}}, \bibinfo {author} {\bibfnamefont {S.~E.}\ \bibnamefont {Economou}}, \bibinfo {author} {\bibfnamefont {c.~C.}\ \bibnamefont {B\u{a}descu}}, \bibinfo {author} {\bibfnamefont {M.}~\bibnamefont {Scheibner}}, \bibinfo {author} {\bibfnamefont {A.~S.}\ \bibnamefont {Bracker}}, \bibinfo {author} {\bibfnamefont {M.}~\bibnamefont {Bashkansky}}, \bibinfo {author} {\bibfnamefont {T.~L.}\ \bibnamefont {Reinecke}},\ and\ \bibinfo {author} {\bibfnamefont {D.}~\bibnamefont {Gammon}},\ }\bibfield  {title} {\bibinfo {title} {Optical spin initialization and non-destructive measurement in a quantum dot molecule},\ }\href {https://doi.org/10.1103/PhysRevLett.101.236804} {\bibfield  {journal} {\bibinfo  {journal} {Physical Review Letters}\ }\textbf {\bibinfo {volume} {101}},\ \bibinfo {pages} {236804} (\bibinfo {year} {2008})}\BibitemShut {NoStop}%
\bibitem [{\citenamefont {Wigger}\ \emph {et~al.}(2023)\citenamefont {Wigger}, \citenamefont {Schall}, \citenamefont {Deconinck}, \citenamefont {Bart}, \citenamefont {Mrowiński}, \citenamefont {Krzykowski}, \citenamefont {Gawarecki}, \citenamefont {von Helversen}, \citenamefont {Schmidt}, \citenamefont {Bremer}, \citenamefont {Bopp}, \citenamefont {Reuter}, \citenamefont {Wieck}, \citenamefont {Rodt}, \citenamefont {Renard}, \citenamefont {Nogues}, \citenamefont {Ludwig}, \citenamefont {Machnikowski}, \citenamefont {Finley}, \citenamefont {Reitzenstein},\ and\ \citenamefont {Kasprzak}}]{wigger_2023}%
  \BibitemOpen
  \bibfield  {author} {\bibinfo {author} {\bibfnamefont {D.}~\bibnamefont {Wigger}}, \bibinfo {author} {\bibfnamefont {J.}~\bibnamefont {Schall}}, \bibinfo {author} {\bibfnamefont {M.}~\bibnamefont {Deconinck}}, \bibinfo {author} {\bibfnamefont {N.}~\bibnamefont {Bart}}, \bibinfo {author} {\bibfnamefont {P.}~\bibnamefont {Mrowiński}}, \bibinfo {author} {\bibfnamefont {M.}~\bibnamefont {Krzykowski}}, \bibinfo {author} {\bibfnamefont {K.}~\bibnamefont {Gawarecki}}, \bibinfo {author} {\bibfnamefont {M.}~\bibnamefont {von Helversen}}, \bibinfo {author} {\bibfnamefont {R.}~\bibnamefont {Schmidt}}, \bibinfo {author} {\bibfnamefont {L.}~\bibnamefont {Bremer}}, \bibinfo {author} {\bibfnamefont {F.}~\bibnamefont {Bopp}}, \bibinfo {author} {\bibfnamefont {D.}~\bibnamefont {Reuter}}, \bibinfo {author} {\bibfnamefont {A.~D.}\ \bibnamefont {Wieck}}, \bibinfo {author} {\bibfnamefont {S.}~\bibnamefont {Rodt}}, \bibinfo {author} {\bibfnamefont {J.}~\bibnamefont {Renard}}, \bibinfo {author} {\bibfnamefont {G.}~\bibnamefont
  {Nogues}}, \bibinfo {author} {\bibfnamefont {A.}~\bibnamefont {Ludwig}}, \bibinfo {author} {\bibfnamefont {P.}~\bibnamefont {Machnikowski}}, \bibinfo {author} {\bibfnamefont {J.~J.}\ \bibnamefont {Finley}}, \bibinfo {author} {\bibfnamefont {S.}~\bibnamefont {Reitzenstein}},\ and\ \bibinfo {author} {\bibfnamefont {J.}~\bibnamefont {Kasprzak}},\ }\href {https://arxiv.org/abs/2304.10148} {\bibinfo {title} {Controlled coherent coupling in a quantum dot molecule revealed by ultrafast four-wave mixing spectroscopy}} (\bibinfo {year} {2023}),\ \Eprint {https://arxiv.org/abs/2304.10148} {arXiv:2304.10148 [cond-mat.mes-hall]} \BibitemShut {NoStop}%
\bibitem [{\citenamefont {Yugova}\ \emph {et~al.}(2009)\citenamefont {Yugova}, \citenamefont {Glazov}, \citenamefont {Ivchenko},\ and\ \citenamefont {Efros}}]{Yugova_2009}%
  \BibitemOpen
  \bibfield  {author} {\bibinfo {author} {\bibfnamefont {I.~A.}\ \bibnamefont {Yugova}}, \bibinfo {author} {\bibfnamefont {M.~M.}\ \bibnamefont {Glazov}}, \bibinfo {author} {\bibfnamefont {E.~L.}\ \bibnamefont {Ivchenko}},\ and\ \bibinfo {author} {\bibfnamefont {A.~L.}\ \bibnamefont {Efros}},\ }\bibfield  {title} {\bibinfo {title} {Pump-probe faraday rotation and ellipticity in an ensemble of singly charged quantum dots},\ }\href {https://doi.org/10.1103/PhysRevB.80.104436} {\bibfield  {journal} {\bibinfo  {journal} {Phys. Rev. B}\ }\textbf {\bibinfo {volume} {80}},\ \bibinfo {pages} {104436} (\bibinfo {year} {2009})}\BibitemShut {NoStop}%
\bibitem [{\citenamefont {Prechtel}\ \emph {et~al.}(2014)\citenamefont {Prechtel}, \citenamefont {Maier}, \citenamefont {Houel}, \citenamefont {Kuhlmann}, \citenamefont {Ludwig}, \citenamefont {Wieck}, \citenamefont {Loss},\ and\ \citenamefont {Warburton}}]{Prechtel_2014}%
  \BibitemOpen
  \bibfield  {author} {\bibinfo {author} {\bibfnamefont {J.~H.}\ \bibnamefont {Prechtel}}, \bibinfo {author} {\bibfnamefont {F.~M.}\ \bibnamefont {Maier}}, \bibinfo {author} {\bibfnamefont {J.}~\bibnamefont {Houel}}, \bibinfo {author} {\bibfnamefont {A.~V.}\ \bibnamefont {Kuhlmann}}, \bibinfo {author} {\bibfnamefont {A.}~\bibnamefont {Ludwig}}, \bibinfo {author} {\bibfnamefont {A.~D.}\ \bibnamefont {Wieck}}, \bibinfo {author} {\bibfnamefont {D.}~\bibnamefont {Loss}},\ and\ \bibinfo {author} {\bibfnamefont {R.~J.}\ \bibnamefont {Warburton}},\ }\bibfield  {title} {\bibinfo {title} {Electrically tunable hole g factor of an optically active quantum dot for fast spin rotations},\ }\href {https://api.semanticscholar.org/CorpusID:119208791} {\bibfield  {journal} {\bibinfo  {journal} {Physical Review B}\ }\textbf {\bibinfo {volume} {91}},\ \bibinfo {pages} {165304} (\bibinfo {year} {2014})}\BibitemShut {NoStop}%
\bibitem [{\citenamefont {Sapienza}\ \emph {et~al.}(2016)\citenamefont {Sapienza}, \citenamefont {Al-Khuzheyri}, \citenamefont {Dada}, \citenamefont {Griffiths}, \citenamefont {Clarke},\ and\ \citenamefont {Gerardot}}]{Sapienza_2016}%
  \BibitemOpen
  \bibfield  {author} {\bibinfo {author} {\bibfnamefont {L.}~\bibnamefont {Sapienza}}, \bibinfo {author} {\bibfnamefont {R.~M.}\ \bibnamefont {Al-Khuzheyri}}, \bibinfo {author} {\bibfnamefont {A.~C.}\ \bibnamefont {Dada}}, \bibinfo {author} {\bibfnamefont {A.}~\bibnamefont {Griffiths}}, \bibinfo {author} {\bibfnamefont {E.}~\bibnamefont {Clarke}},\ and\ \bibinfo {author} {\bibfnamefont {B.~D.}\ \bibnamefont {Gerardot}},\ }\bibfield  {title} {\bibinfo {title} {Magneto-optical spectroscopy of single charge-tunable {InAs/GaAs} quantum dots emitting at telecom wavelengths},\ }\href {https://api.semanticscholar.org/CorpusID:8554196} {\bibfield  {journal} {\bibinfo  {journal} {Physical Review B}\ }\textbf {\bibinfo {volume} {93}},\ \bibinfo {pages} {155301} (\bibinfo {year} {2016})}\BibitemShut {NoStop}%
\bibitem [{\citenamefont {Belykh}\ \emph {et~al.}(2016{\natexlab{a}})\citenamefont {Belykh}, \citenamefont {Yakovlev}, \citenamefont {Schindler}, \citenamefont {van Bree}, \citenamefont {Koenraad}, \citenamefont {Averkiev}, \citenamefont {Bayer},\ and\ \citenamefont {Silov}}]{Belykh_2016}%
  \BibitemOpen
  \bibfield  {author} {\bibinfo {author} {\bibfnamefont {V.~V.}\ \bibnamefont {Belykh}}, \bibinfo {author} {\bibfnamefont {D.~R.}\ \bibnamefont {Yakovlev}}, \bibinfo {author} {\bibfnamefont {J.~J.}\ \bibnamefont {Schindler}}, \bibinfo {author} {\bibfnamefont {J.}~\bibnamefont {van Bree}}, \bibinfo {author} {\bibfnamefont {P.~M.}\ \bibnamefont {Koenraad}}, \bibinfo {author} {\bibfnamefont {N.~S.}\ \bibnamefont {Averkiev}}, \bibinfo {author} {\bibfnamefont {M.}~\bibnamefont {Bayer}},\ and\ \bibinfo {author} {\bibfnamefont {A.~Y.}\ \bibnamefont {Silov}},\ }\bibfield  {title} {\bibinfo {title} {Dispersion of the electron g factor anisotropy in {InAs/InP} self-assembled quantum dots},\ }\href {https://doi.org/10.1063/1.4961201} {\bibfield  {journal} {\bibinfo  {journal} {Journal of Applied Physics}\ }\textbf {\bibinfo {volume} {120}},\ \bibinfo {pages} {084301} (\bibinfo {year} {2016}{\natexlab{a}})}\BibitemShut {NoStop}%
\bibitem [{\citenamefont {Kim}\ \emph {et~al.}(2009)\citenamefont {Kim}, \citenamefont {Sheng}, \citenamefont {Poole}, \citenamefont {Dalacu}, \citenamefont {Lefebvre}, \citenamefont {Lapointe}, \citenamefont {Reimer}, \citenamefont {Aers},\ and\ \citenamefont {Williams}}]{Kim_2009}%
  \BibitemOpen
  \bibfield  {author} {\bibinfo {author} {\bibfnamefont {D.}~\bibnamefont {Kim}}, \bibinfo {author} {\bibfnamefont {W.}~\bibnamefont {Sheng}}, \bibinfo {author} {\bibfnamefont {P.~J.}\ \bibnamefont {Poole}}, \bibinfo {author} {\bibfnamefont {D.}~\bibnamefont {Dalacu}}, \bibinfo {author} {\bibfnamefont {J.}~\bibnamefont {Lefebvre}}, \bibinfo {author} {\bibfnamefont {J.}~\bibnamefont {Lapointe}}, \bibinfo {author} {\bibfnamefont {M.~E.}\ \bibnamefont {Reimer}}, \bibinfo {author} {\bibfnamefont {G.~C.}\ \bibnamefont {Aers}},\ and\ \bibinfo {author} {\bibfnamefont {R.~L.}\ \bibnamefont {Williams}},\ }\bibfield  {title} {\bibinfo {title} {Tuning the exciton $g$ factor in single {InAs/InP} quantum dots},\ }\href {https://doi.org/10.1103/PhysRevB.79.045310} {\bibfield  {journal} {\bibinfo  {journal} {Phys. Rev. B}\ }\textbf {\bibinfo {volume} {79}},\ \bibinfo {pages} {045310} (\bibinfo {year} {2009})}\BibitemShut {NoStop}%
\bibitem [{\citenamefont {Belykh}\ \emph {et~al.}(2015)\citenamefont {Belykh}, \citenamefont {Greilich}, \citenamefont {Yakovlev}, \citenamefont {Yacob}, \citenamefont {Reithmaier}, \citenamefont {Benyoucef},\ and\ \citenamefont {Bayer}}]{Belykh_2015}%
  \BibitemOpen
  \bibfield  {author} {\bibinfo {author} {\bibfnamefont {V.~V.}\ \bibnamefont {Belykh}}, \bibinfo {author} {\bibfnamefont {A.}~\bibnamefont {Greilich}}, \bibinfo {author} {\bibfnamefont {D.~R.}\ \bibnamefont {Yakovlev}}, \bibinfo {author} {\bibfnamefont {M.}~\bibnamefont {Yacob}}, \bibinfo {author} {\bibfnamefont {J.~P.}\ \bibnamefont {Reithmaier}}, \bibinfo {author} {\bibfnamefont {M.}~\bibnamefont {Benyoucef}},\ and\ \bibinfo {author} {\bibfnamefont {M.}~\bibnamefont {Bayer}},\ }\bibfield  {title} {\bibinfo {title} {Electron and hole $g$ factors in {InAs/InAlGaAs} self-assembled quantum dots emitting at telecom wavelengths},\ }\href {https://doi.org/10.1103/PhysRevB.92.165307} {\bibfield  {journal} {\bibinfo  {journal} {Phys. Rev. B}\ }\textbf {\bibinfo {volume} {92}},\ \bibinfo {pages} {165307} (\bibinfo {year} {2015})}\BibitemShut {NoStop}%
\bibitem [{\citenamefont {Zhukov}\ \emph {et~al.}(2018)\citenamefont {Zhukov}, \citenamefont {Kirstein}, \citenamefont {Smirnov}, \citenamefont {Yakovlev}, \citenamefont {Glazov}, \citenamefont {Reuter}, \citenamefont {Wieck}, \citenamefont {Bayer},\ and\ \citenamefont {Greilich}}]{Zhukov_2018}%
  \BibitemOpen
  \bibfield  {author} {\bibinfo {author} {\bibfnamefont {E.~A.}\ \bibnamefont {Zhukov}}, \bibinfo {author} {\bibfnamefont {E.}~\bibnamefont {Kirstein}}, \bibinfo {author} {\bibfnamefont {D.~S.}\ \bibnamefont {Smirnov}}, \bibinfo {author} {\bibfnamefont {D.~R.}\ \bibnamefont {Yakovlev}}, \bibinfo {author} {\bibfnamefont {M.~M.}\ \bibnamefont {Glazov}}, \bibinfo {author} {\bibfnamefont {D.}~\bibnamefont {Reuter}}, \bibinfo {author} {\bibfnamefont {A.~D.}\ \bibnamefont {Wieck}}, \bibinfo {author} {\bibfnamefont {M.}~\bibnamefont {Bayer}},\ and\ \bibinfo {author} {\bibfnamefont {A.}~\bibnamefont {Greilich}},\ }\bibfield  {title} {\bibinfo {title} {Spin inertia of resident and photoexcited carriers in singly charged quantum dots},\ }\href {https://doi.org/10.1103/PhysRevB.98.121304} {\bibfield  {journal} {\bibinfo  {journal} {Phys. Rev. B}\ }\textbf {\bibinfo {volume} {98}},\ \bibinfo {pages} {121304} (\bibinfo {year} {2018})}\BibitemShut {NoStop}%
\bibitem [{\citenamefont {Mikhailov}\ \emph {et~al.}(2008)\citenamefont {Mikhailov}, \citenamefont {García},\ and\ \citenamefont {Marín}}]{Mikhailov_2008}%
  \BibitemOpen
  \bibfield  {author} {\bibinfo {author} {\bibfnamefont {I.}~\bibnamefont {Mikhailov}}, \bibinfo {author} {\bibfnamefont {L.}~\bibnamefont {García}},\ and\ \bibinfo {author} {\bibfnamefont {J.}~\bibnamefont {Marín}},\ }\bibfield  {title} {\bibinfo {title} {Vertically coupled quantum dots charged by exciton},\ }\href {https://doi.org/https://doi.org/10.1016/j.mejo.2007.07.046} {\bibfield  {journal} {\bibinfo  {journal} {Microelectronics Journal}\ }\textbf {\bibinfo {volume} {39}},\ \bibinfo {pages} {378} (\bibinfo {year} {2008})},\ \bibinfo {note} {the Sixth International Conference on Low Dimensional Structures and Devices}\BibitemShut {NoStop}%
\bibitem [{\citenamefont {Testelin}\ \emph {et~al.}(2009)\citenamefont {Testelin}, \citenamefont {Bernardot}, \citenamefont {Eble},\ and\ \citenamefont {Chamarro}}]{Testelin_2009}%
  \BibitemOpen
  \bibfield  {author} {\bibinfo {author} {\bibfnamefont {C.}~\bibnamefont {Testelin}}, \bibinfo {author} {\bibfnamefont {F.}~\bibnamefont {Bernardot}}, \bibinfo {author} {\bibfnamefont {B.}~\bibnamefont {Eble}},\ and\ \bibinfo {author} {\bibfnamefont {M.}~\bibnamefont {Chamarro}},\ }\bibfield  {title} {\bibinfo {title} {Hole--spin dephasing time associated with hyperfine interaction in quantum dots},\ }\href {https://doi.org/10.1103/PhysRevB.79.195440} {\bibfield  {journal} {\bibinfo  {journal} {Phys. Rev. B}\ }\textbf {\bibinfo {volume} {79}},\ \bibinfo {pages} {195440} (\bibinfo {year} {2009})}\BibitemShut {NoStop}%
\bibitem [{\citenamefont {Cizauskas}\ \emph {et~al.}(2025)\citenamefont {Cizauskas}, \citenamefont {Sala}, \citenamefont {Heffernan}, \citenamefont {Fox}, \citenamefont {Bayer},\ and\ \citenamefont {Greilich}}]{Marius_2025}%
  \BibitemOpen
  \bibfield  {author} {\bibinfo {author} {\bibfnamefont {M.}~\bibnamefont {Cizauskas}}, \bibinfo {author} {\bibfnamefont {E.~M.}\ \bibnamefont {Sala}}, \bibinfo {author} {\bibfnamefont {J.}~\bibnamefont {Heffernan}}, \bibinfo {author} {\bibfnamefont {A.~M.}\ \bibnamefont {Fox}}, \bibinfo {author} {\bibfnamefont {M.}~\bibnamefont {Bayer}},\ and\ \bibinfo {author} {\bibfnamefont {A.}~\bibnamefont {Greilich}},\ }\bibfield  {title} {\bibinfo {title} {Spin properties in droplet epitaxy-grown telecom quantum dots},\ }\href@noop {} {\bibfield  {journal} {\bibinfo  {journal} {Arxiv}\ } (\bibinfo {year} {2025})}\BibitemShut {NoStop}%
\bibitem [{\citenamefont {Hostein}\ \emph {et~al.}(2008)\citenamefont {Hostein}, \citenamefont {Michon}, \citenamefont {Beaudoin}, \citenamefont {Gogneau}, \citenamefont {Patriache}, \citenamefont {Marzin}, \citenamefont {Robert-Philip}, \citenamefont {Sagnes},\ and\ \citenamefont {Beveratos}}]{Hostein_2008}%
  \BibitemOpen
  \bibfield  {author} {\bibinfo {author} {\bibfnamefont {R.}~\bibnamefont {Hostein}}, \bibinfo {author} {\bibfnamefont {A.}~\bibnamefont {Michon}}, \bibinfo {author} {\bibfnamefont {G.}~\bibnamefont {Beaudoin}}, \bibinfo {author} {\bibfnamefont {N.}~\bibnamefont {Gogneau}}, \bibinfo {author} {\bibfnamefont {G.}~\bibnamefont {Patriache}}, \bibinfo {author} {\bibfnamefont {J.-Y.}\ \bibnamefont {Marzin}}, \bibinfo {author} {\bibfnamefont {I.}~\bibnamefont {Robert-Philip}}, \bibinfo {author} {\bibfnamefont {I.}~\bibnamefont {Sagnes}},\ and\ \bibinfo {author} {\bibfnamefont {A.}~\bibnamefont {Beveratos}},\ }\bibfield  {title} {\bibinfo {title} {Time-resolved characterization of {InAsP/InP} quantum dots emitting in the c-band telecommunication window},\ }\href {https://doi.org/10.1063/1.2965112} {\bibfield  {journal} {\bibinfo  {journal} {Applied Physics Letters}\ }\textbf {\bibinfo {volume} {93}},\ \bibinfo {pages} {073106} (\bibinfo {year} {2008})},\ \Eprint
  {https://arxiv.org/abs/https://pubs.aip.org/aip/apl/article-pdf/doi/10.1063/1.2965112/13710102/073106\_1\_online.pdf} {https://pubs.aip.org/aip/apl/article-pdf/doi/10.1063/1.2965112/13710102/073106\_1\_online.pdf} \BibitemShut {NoStop}%
\bibitem [{\citenamefont {Petrov}\ \emph {et~al.}(2008)\citenamefont {Petrov}, \citenamefont {Ignatiev}, \citenamefont {Poltavtsev}, \citenamefont {Greilich}, \citenamefont {Bauschulte}, \citenamefont {Yakovlev},\ and\ \citenamefont {Bayer}}]{Petrov_2008}%
  \BibitemOpen
  \bibfield  {author} {\bibinfo {author} {\bibfnamefont {M.~Y.}\ \bibnamefont {Petrov}}, \bibinfo {author} {\bibfnamefont {I.~V.}\ \bibnamefont {Ignatiev}}, \bibinfo {author} {\bibfnamefont {S.~V.}\ \bibnamefont {Poltavtsev}}, \bibinfo {author} {\bibfnamefont {A.}~\bibnamefont {Greilich}}, \bibinfo {author} {\bibfnamefont {A.}~\bibnamefont {Bauschulte}}, \bibinfo {author} {\bibfnamefont {D.~R.}\ \bibnamefont {Yakovlev}},\ and\ \bibinfo {author} {\bibfnamefont {M.}~\bibnamefont {Bayer}},\ }\bibfield  {title} {\bibinfo {title} {Effect of thermal annealing on the hyperfine interaction in inas/gaas quantum dots},\ }\href {https://doi.org/10.1103/PhysRevB.78.045315} {\bibfield  {journal} {\bibinfo  {journal} {Phys. Rev. B}\ }\textbf {\bibinfo {volume} {78}},\ \bibinfo {pages} {045315} (\bibinfo {year} {2008})}\BibitemShut {NoStop}%
\bibitem [{\citenamefont {Greilich}\ \emph {et~al.}(2012)\citenamefont {Greilich}, \citenamefont {Pawlis}, \citenamefont {Liu}, \citenamefont {Yugov}, \citenamefont {Yakovlev}, \citenamefont {Lischka}, \citenamefont {Yamamoto},\ and\ \citenamefont {Bayer}}]{Greilich_2012}%
  \BibitemOpen
  \bibfield  {author} {\bibinfo {author} {\bibfnamefont {A.}~\bibnamefont {Greilich}}, \bibinfo {author} {\bibfnamefont {A.}~\bibnamefont {Pawlis}}, \bibinfo {author} {\bibfnamefont {F.}~\bibnamefont {Liu}}, \bibinfo {author} {\bibfnamefont {O.~A.}\ \bibnamefont {Yugov}}, \bibinfo {author} {\bibfnamefont {D.~R.}\ \bibnamefont {Yakovlev}}, \bibinfo {author} {\bibfnamefont {K.}~\bibnamefont {Lischka}}, \bibinfo {author} {\bibfnamefont {Y.}~\bibnamefont {Yamamoto}},\ and\ \bibinfo {author} {\bibfnamefont {M.}~\bibnamefont {Bayer}},\ }\bibfield  {title} {\bibinfo {title} {Spin dephasing of fluorine-bound electrons in {ZnSe}},\ }\href {https://doi.org/10.1103/PhysRevB.85.121303} {\bibfield  {journal} {\bibinfo  {journal} {Phys. Rev. B}\ }\textbf {\bibinfo {volume} {85}},\ \bibinfo {pages} {121303} (\bibinfo {year} {2012})}\BibitemShut {NoStop}%
\bibitem [{\citenamefont {Stockill}\ \emph {et~al.}(2016)\citenamefont {Stockill}, \citenamefont {Le~Gall}, \citenamefont {Matthiesen}, \citenamefont {Huthmacher}, \citenamefont {Clarke}, \citenamefont {Hugues},\ and\ \citenamefont {Atat{\"{u}}re}}]{Stockill_2016}%
  \BibitemOpen
  \bibfield  {author} {\bibinfo {author} {\bibfnamefont {R.}~\bibnamefont {Stockill}}, \bibinfo {author} {\bibfnamefont {C.}~\bibnamefont {Le~Gall}}, \bibinfo {author} {\bibfnamefont {C.}~\bibnamefont {Matthiesen}}, \bibinfo {author} {\bibfnamefont {L.}~\bibnamefont {Huthmacher}}, \bibinfo {author} {\bibfnamefont {E.}~\bibnamefont {Clarke}}, \bibinfo {author} {\bibfnamefont {M.}~\bibnamefont {Hugues}},\ and\ \bibinfo {author} {\bibfnamefont {M.}~\bibnamefont {Atat{\"{u}}re}},\ }\bibfield  {title} {\bibinfo {title} {Quantum dot spin coherence governed by a strained nuclear environment},\ }\href {https://doi.org/10.1038/ncomms12745} {\bibfield  {journal} {\bibinfo  {journal} {Nature Communications}\ }\textbf {\bibinfo {volume} {7}},\ \bibinfo {pages} {12745} (\bibinfo {year} {2016})}\BibitemShut {NoStop}%
\bibitem [{\citenamefont {Godden}\ \emph {et~al.}(2012)\citenamefont {Godden}, \citenamefont {Quilter}, \citenamefont {Ramsay}, \citenamefont {Wu}, \citenamefont {Brereton}, \citenamefont {Boyle}, \citenamefont {Luxmoore}, \citenamefont {Heberle}, \citenamefont {Skolnick},\ and\ \citenamefont {Fox}}]{Godden2012}%
  \BibitemOpen
  \bibfield  {author} {\bibinfo {author} {\bibfnamefont {T.~M.}\ \bibnamefont {Godden}}, \bibinfo {author} {\bibfnamefont {J.~H.}\ \bibnamefont {Quilter}}, \bibinfo {author} {\bibfnamefont {A.~J.}\ \bibnamefont {Ramsay}}, \bibinfo {author} {\bibfnamefont {Y.}~\bibnamefont {Wu}}, \bibinfo {author} {\bibfnamefont {P.}~\bibnamefont {Brereton}}, \bibinfo {author} {\bibfnamefont {S.~J.}\ \bibnamefont {Boyle}}, \bibinfo {author} {\bibfnamefont {I.~J.}\ \bibnamefont {Luxmoore}}, \bibinfo {author} {\bibfnamefont {A.~P.}\ \bibnamefont {Heberle}}, \bibinfo {author} {\bibfnamefont {M.~S.}\ \bibnamefont {Skolnick}},\ and\ \bibinfo {author} {\bibfnamefont {A.~M.}\ \bibnamefont {Fox}},\ }\bibfield  {title} {\bibinfo {title} {Coherent optical control of the spin of a single hole in an {InAs/GaAs} quantum dot},\ }\href {https://doi.org/10.1103/PhysRevLett.108.017402} {\bibfield  {journal} {\bibinfo  {journal} {Phys. Rev. Lett.}\ }\textbf {\bibinfo {volume} {108}},\ \bibinfo {pages} {017402} (\bibinfo {year}
  {2012})}\BibitemShut {NoStop}%
\bibitem [{\citenamefont {Heisterkamp}\ \emph {et~al.}(2015)\citenamefont {Heisterkamp}, \citenamefont {Zhukov}, \citenamefont {Greilich}, \citenamefont {Yakovlev}, \citenamefont {Korenev}, \citenamefont {Pawlis},\ and\ \citenamefont {Bayer}}]{Heisterkamp_2015}%
  \BibitemOpen
  \bibfield  {author} {\bibinfo {author} {\bibfnamefont {F.}~\bibnamefont {Heisterkamp}}, \bibinfo {author} {\bibfnamefont {E.~A.}\ \bibnamefont {Zhukov}}, \bibinfo {author} {\bibfnamefont {A.}~\bibnamefont {Greilich}}, \bibinfo {author} {\bibfnamefont {D.~R.}\ \bibnamefont {Yakovlev}}, \bibinfo {author} {\bibfnamefont {V.~L.}\ \bibnamefont {Korenev}}, \bibinfo {author} {\bibfnamefont {A.}~\bibnamefont {Pawlis}},\ and\ \bibinfo {author} {\bibfnamefont {M.}~\bibnamefont {Bayer}},\ }\bibfield  {title} {\bibinfo {title} {Longitudinal and transverse spin dynamics of donor-bound electrons in fluorine-doped znse: Spin inertia versus hanle effect},\ }\href {https://doi.org/10.1103/PhysRevB.91.235432} {\bibfield  {journal} {\bibinfo  {journal} {Phys. Rev. B}\ }\textbf {\bibinfo {volume} {91}},\ \bibinfo {pages} {235432} (\bibinfo {year} {2015})}\BibitemShut {NoStop}%
\bibitem [{\citenamefont {Bulaev}\ and\ \citenamefont {Loss}(2005)}]{Bulaev_2005}%
  \BibitemOpen
  \bibfield  {author} {\bibinfo {author} {\bibfnamefont {D.~V.}\ \bibnamefont {Bulaev}}\ and\ \bibinfo {author} {\bibfnamefont {D.}~\bibnamefont {Loss}},\ }\bibfield  {title} {\bibinfo {title} {Spin relaxation and decoherence of holes in quantum dots},\ }\href {https://doi.org/10.1103/PhysRevLett.95.076805} {\bibfield  {journal} {\bibinfo  {journal} {Phys. Rev. Lett.}\ }\textbf {\bibinfo {volume} {95}},\ \bibinfo {pages} {076805} (\bibinfo {year} {2005})}\BibitemShut {NoStop}%
\bibitem [{\citenamefont {Chirolli}\ and\ \citenamefont {Burkard}(2008)}]{Chirolli_2008}%
  \BibitemOpen
  \bibfield  {author} {\bibinfo {author} {\bibfnamefont {L.}~\bibnamefont {Chirolli}}\ and\ \bibinfo {author} {\bibfnamefont {G.}~\bibnamefont {Burkard}},\ }\bibfield  {title} {\bibinfo {title} {Decoherence in solid-state qubits},\ }\href {https://api.semanticscholar.org/CorpusID:9025611} {\bibfield  {journal} {\bibinfo  {journal} {Advances in Physics}\ }\textbf {\bibinfo {volume} {57}},\ \bibinfo {pages} {225 } (\bibinfo {year} {2008})}\BibitemShut {NoStop}%
\bibitem [{\citenamefont {Wei}\ \emph {et~al.}(2010)\citenamefont {Wei}, \citenamefont {Gong}, \citenamefont {Guo},\ and\ \citenamefont {He}}]{Wei_2010}%
  \BibitemOpen
  \bibfield  {author} {\bibinfo {author} {\bibfnamefont {H.}~\bibnamefont {Wei}}, \bibinfo {author} {\bibfnamefont {M.}~\bibnamefont {Gong}}, \bibinfo {author} {\bibfnamefont {G.}~\bibnamefont {Guo}},\ and\ \bibinfo {author} {\bibfnamefont {L.}~\bibnamefont {He}},\ }\bibfield  {title} {\bibinfo {title} {Atomistic pseudopotential theory of spin relaxation in self-assembled {In1-xGaxAs/GaAs} quantum dots at zero magnetic field},\ }\href {https://api.semanticscholar.org/CorpusID:118382071} {\bibfield  {journal} {\bibinfo  {journal} {Physical Review B}\ }\textbf {\bibinfo {volume} {85}},\ \bibinfo {pages} {045317} (\bibinfo {year} {2010})}\BibitemShut {NoStop}%
\bibitem [{\citenamefont {Zhukov}\ \emph {et~al.}(2007)\citenamefont {Zhukov}, \citenamefont {Yakovlev}, \citenamefont {Bayer}, \citenamefont {Glazov}, \citenamefont {Ivchenko}, \citenamefont {Karczewski}, \citenamefont {Wojtowicz},\ and\ \citenamefont {Kossut}}]{Zhukov_2007}%
  \BibitemOpen
  \bibfield  {author} {\bibinfo {author} {\bibfnamefont {E.~A.}\ \bibnamefont {Zhukov}}, \bibinfo {author} {\bibfnamefont {D.~R.}\ \bibnamefont {Yakovlev}}, \bibinfo {author} {\bibfnamefont {M.}~\bibnamefont {Bayer}}, \bibinfo {author} {\bibfnamefont {M.~M.}\ \bibnamefont {Glazov}}, \bibinfo {author} {\bibfnamefont {E.~L.}\ \bibnamefont {Ivchenko}}, \bibinfo {author} {\bibfnamefont {G.}~\bibnamefont {Karczewski}}, \bibinfo {author} {\bibfnamefont {T.}~\bibnamefont {Wojtowicz}},\ and\ \bibinfo {author} {\bibfnamefont {J.}~\bibnamefont {Kossut}},\ }\bibfield  {title} {\bibinfo {title} {Spin coherence of a two-dimensional electron gas induced by resonant excitation of trions and excitons in {CdTe/(Cd,Mg)Te} quantum wells},\ }\href {https://api.semanticscholar.org/CorpusID:119317616} {\bibfield  {journal} {\bibinfo  {journal} {Physical Review B}\ }\textbf {\bibinfo {volume} {76}},\ \bibinfo {pages} {205310} (\bibinfo {year} {2007})}\BibitemShut {NoStop}%
\bibitem [{\citenamefont {Greilich}\ \emph {et~al.}(2006)\citenamefont {Greilich}, \citenamefont {Yakovlev}, \citenamefont {Shabaev}, \citenamefont {Efros}, \citenamefont {Yugova}, \citenamefont {Oulton}, \citenamefont {Stavarache}, \citenamefont {Reuter}, \citenamefont {Wieck},\ and\ \citenamefont {Bayer}}]{Greilich_2006}%
  \BibitemOpen
  \bibfield  {author} {\bibinfo {author} {\bibfnamefont {A.}~\bibnamefont {Greilich}}, \bibinfo {author} {\bibfnamefont {D.~R.}\ \bibnamefont {Yakovlev}}, \bibinfo {author} {\bibfnamefont {A.}~\bibnamefont {Shabaev}}, \bibinfo {author} {\bibfnamefont {A.~L.}\ \bibnamefont {Efros}}, \bibinfo {author} {\bibfnamefont {I.~A.}\ \bibnamefont {Yugova}}, \bibinfo {author} {\bibfnamefont {R.}~\bibnamefont {Oulton}}, \bibinfo {author} {\bibfnamefont {V.}~\bibnamefont {Stavarache}}, \bibinfo {author} {\bibfnamefont {D.}~\bibnamefont {Reuter}}, \bibinfo {author} {\bibfnamefont {A.}~\bibnamefont {Wieck}},\ and\ \bibinfo {author} {\bibfnamefont {M.}~\bibnamefont {Bayer}},\ }\bibfield  {title} {\bibinfo {title} {Mode locking of electron spin coherences in singly charged quantum dots},\ }\href {https://doi.org/10.1126/science.1128215} {\bibfield  {journal} {\bibinfo  {journal} {Science}\ }\textbf {\bibinfo {volume} {313}},\ \bibinfo {pages} {341} (\bibinfo {year} {2006})}\BibitemShut {NoStop}%
\bibitem [{\citenamefont {Yugova}\ \emph {et~al.}(2012)\citenamefont {Yugova}, \citenamefont {Glazov}, \citenamefont {Yakovlev}, \citenamefont {Sokolova},\ and\ \citenamefont {Bayer}}]{Yugova_2012}%
  \BibitemOpen
  \bibfield  {author} {\bibinfo {author} {\bibfnamefont {I.}~\bibnamefont {Yugova}}, \bibinfo {author} {\bibfnamefont {M.}~\bibnamefont {Glazov}}, \bibinfo {author} {\bibfnamefont {D.}~\bibnamefont {Yakovlev}}, \bibinfo {author} {\bibfnamefont {A.}~\bibnamefont {Sokolova}},\ and\ \bibinfo {author} {\bibfnamefont {M.}~\bibnamefont {Bayer}},\ }\bibfield  {title} {\bibinfo {title} {Coherent spin dynamics of electrons and holes in semiconductor quantum wells and quantum dots under periodical optical excitation: Resonant spin amplification versus spin mode locking},\ }\href@noop {} {\bibfield  {journal} {\bibinfo  {journal} {Physical Review B—Condensed Matter and Materials Physics}\ }\textbf {\bibinfo {volume} {85}},\ \bibinfo {pages} {125304} (\bibinfo {year} {2012})}\BibitemShut {NoStop}%
\bibitem [{\citenamefont {Krenner}\ \emph {et~al.}(2005)\citenamefont {Krenner}, \citenamefont {Stufler}, \citenamefont {Sabathil}, \citenamefont {Clark}, \citenamefont {Ester}, \citenamefont {Bichler}, \citenamefont {Abstreiter}, \citenamefont {Finley},\ and\ \citenamefont {Zrenner}}]{Krenner_2005}%
  \BibitemOpen
  \bibfield  {author} {\bibinfo {author} {\bibfnamefont {H.~J.}\ \bibnamefont {Krenner}}, \bibinfo {author} {\bibfnamefont {S.}~\bibnamefont {Stufler}}, \bibinfo {author} {\bibfnamefont {M.}~\bibnamefont {Sabathil}}, \bibinfo {author} {\bibfnamefont {E.~C.}\ \bibnamefont {Clark}}, \bibinfo {author} {\bibfnamefont {P.}~\bibnamefont {Ester}}, \bibinfo {author} {\bibfnamefont {M.}~\bibnamefont {Bichler}}, \bibinfo {author} {\bibfnamefont {G.}~\bibnamefont {Abstreiter}}, \bibinfo {author} {\bibfnamefont {J.~J.}\ \bibnamefont {Finley}},\ and\ \bibinfo {author} {\bibfnamefont {A.}~\bibnamefont {Zrenner}},\ }\bibfield  {title} {\bibinfo {title} {Recent advances in exciton-based quantum information processing in quantum dot nanostructures},\ }\href {https://api.semanticscholar.org/CorpusID:119384661} {\bibfield  {journal} {\bibinfo  {journal} {New Journal of Physics}\ }\textbf {\bibinfo {volume} {7}},\ \bibinfo {pages} {184 } (\bibinfo {year} {2005})}\BibitemShut {NoStop}%
\bibitem [{\citenamefont {Usman}\ \emph {et~al.}(2011)\citenamefont {Usman}, \citenamefont {Inoue}, \citenamefont {Harda}, \citenamefont {Klimeck},\ and\ \citenamefont {Kita}}]{Usman_2011}%
  \BibitemOpen
  \bibfield  {author} {\bibinfo {author} {\bibfnamefont {M.}~\bibnamefont {Usman}}, \bibinfo {author} {\bibfnamefont {T.}~\bibnamefont {Inoue}}, \bibinfo {author} {\bibfnamefont {Y.}~\bibnamefont {Harda}}, \bibinfo {author} {\bibfnamefont {G.}~\bibnamefont {Klimeck}},\ and\ \bibinfo {author} {\bibfnamefont {T.}~\bibnamefont {Kita}},\ }\bibfield  {title} {\bibinfo {title} {Experimental and atomistic theoretical study of degree of polarization from multilayer {InAs/GaAs} quantum dot stacks},\ }\href {https://api.semanticscholar.org/CorpusID:12600111} {\bibfield  {journal} {\bibinfo  {journal} {Physical Review B}\ }\textbf {\bibinfo {volume} {84}},\ \bibinfo {pages} {115321} (\bibinfo {year} {2011})}\BibitemShut {NoStop}%
\bibitem [{\citenamefont {Luo}\ \emph {et~al.}(2014)\citenamefont {Luo}, \citenamefont {Bester},\ and\ \citenamefont {Zunger}}]{Luo_2014}%
  \BibitemOpen
  \bibfield  {author} {\bibinfo {author} {\bibfnamefont {J.}~\bibnamefont {Luo}}, \bibinfo {author} {\bibfnamefont {G.}~\bibnamefont {Bester}},\ and\ \bibinfo {author} {\bibfnamefont {A.}~\bibnamefont {Zunger}},\ }\bibfield  {title} {\bibinfo {title} {Supercoupling between heavy-hole and light-hole states in nanostructures},\ }\href {https://api.semanticscholar.org/CorpusID:115146423} {\bibfield  {journal} {\bibinfo  {journal} {Physical Review B}\ }\textbf {\bibinfo {volume} {92}} (\bibinfo {year} {2014})}\BibitemShut {NoStop}%
\bibitem [{\citenamefont {Belykh}\ \emph {et~al.}(2016{\natexlab{b}})\citenamefont {Belykh}, \citenamefont {Yakovlev}, \citenamefont {Schindler}, \citenamefont {Zhukov}, \citenamefont {Semina}, \citenamefont {Yacob}, \citenamefont {Reithmaier}, \citenamefont {Benyoucef},\ and\ \citenamefont {Bayer}}]{Belykh_2016_anisotropy}%
  \BibitemOpen
  \bibfield  {author} {\bibinfo {author} {\bibfnamefont {V.~V.}\ \bibnamefont {Belykh}}, \bibinfo {author} {\bibfnamefont {D.~R.}\ \bibnamefont {Yakovlev}}, \bibinfo {author} {\bibfnamefont {J.~J.}\ \bibnamefont {Schindler}}, \bibinfo {author} {\bibfnamefont {E.~A.}\ \bibnamefont {Zhukov}}, \bibinfo {author} {\bibfnamefont {M.~A.}\ \bibnamefont {Semina}}, \bibinfo {author} {\bibfnamefont {M.}~\bibnamefont {Yacob}}, \bibinfo {author} {\bibfnamefont {J.~P.}\ \bibnamefont {Reithmaier}}, \bibinfo {author} {\bibfnamefont {M.}~\bibnamefont {Benyoucef}},\ and\ \bibinfo {author} {\bibfnamefont {M.}~\bibnamefont {Bayer}},\ }\bibfield  {title} {\bibinfo {title} {Large anisotropy of electron and hole $g$ factors in infrared-emitting {InAs/InAlGaAs} self-assembled quantum dots},\ }\href {https://doi.org/10.1103/PhysRevB.93.125302} {\bibfield  {journal} {\bibinfo  {journal} {Phys. Rev. B}\ }\textbf {\bibinfo {volume} {93}},\ \bibinfo {pages} {125302} (\bibinfo {year} {2016}{\natexlab{b}})}\BibitemShut {NoStop}%
\bibitem [{\citenamefont {Marie}\ \emph {et~al.}(1999)\citenamefont {Marie}, \citenamefont {Amand}, \citenamefont {Le~Jeune}, \citenamefont {Paillard}, \citenamefont {Renucci}, \citenamefont {Golub}, \citenamefont {Dymnikov},\ and\ \citenamefont {Ivchenko}}]{Marie_1999}%
  \BibitemOpen
  \bibfield  {author} {\bibinfo {author} {\bibfnamefont {X.}~\bibnamefont {Marie}}, \bibinfo {author} {\bibfnamefont {T.}~\bibnamefont {Amand}}, \bibinfo {author} {\bibfnamefont {P.}~\bibnamefont {Le~Jeune}}, \bibinfo {author} {\bibfnamefont {M.}~\bibnamefont {Paillard}}, \bibinfo {author} {\bibfnamefont {P.}~\bibnamefont {Renucci}}, \bibinfo {author} {\bibfnamefont {L.~E.}\ \bibnamefont {Golub}}, \bibinfo {author} {\bibfnamefont {V.~D.}\ \bibnamefont {Dymnikov}},\ and\ \bibinfo {author} {\bibfnamefont {E.~L.}\ \bibnamefont {Ivchenko}},\ }\bibfield  {title} {\bibinfo {title} {Hole spin quantum beats in quantum-well structures},\ }\href {https://doi.org/10.1103/PhysRevB.60.5811} {\bibfield  {journal} {\bibinfo  {journal} {Phys. Rev. B}\ }\textbf {\bibinfo {volume} {60}},\ \bibinfo {pages} {5811} (\bibinfo {year} {1999})}\BibitemShut {NoStop}%
\bibitem [{\citenamefont {Semina}\ \emph {et~al.}(2023)\citenamefont {Semina}, \citenamefont {Golovatenko},\ and\ \citenamefont {Rodina}}]{Semina_2023}%
  \BibitemOpen
  \bibfield  {author} {\bibinfo {author} {\bibfnamefont {M.~A.}\ \bibnamefont {Semina}}, \bibinfo {author} {\bibfnamefont {A.~A.}\ \bibnamefont {Golovatenko}},\ and\ \bibinfo {author} {\bibfnamefont {A.~V.}\ \bibnamefont {Rodina}},\ }\bibfield  {title} {\bibinfo {title} {Cubic anisotropy of hole zeeman splitting in semiconductor nanocrystals},\ }\href {https://doi.org/10.1103/PhysRevB.108.235310} {\bibfield  {journal} {\bibinfo  {journal} {Phys. Rev. B}\ }\textbf {\bibinfo {volume} {108}},\ \bibinfo {pages} {235310} (\bibinfo {year} {2023})}\BibitemShut {NoStop}%
\bibitem [{\citenamefont {Yu}\ \emph {et~al.}(1996)\citenamefont {Yu}, \citenamefont {Lycett}, \citenamefont {Roberts},\ and\ \citenamefont {Murray}}]{Yu_1996}%
  \BibitemOpen
  \bibfield  {author} {\bibinfo {author} {\bibfnamefont {H.}~\bibnamefont {Yu}}, \bibinfo {author} {\bibfnamefont {S.}~\bibnamefont {Lycett}}, \bibinfo {author} {\bibfnamefont {C.}~\bibnamefont {Roberts}},\ and\ \bibinfo {author} {\bibfnamefont {R.}~\bibnamefont {Murray}},\ }\bibfield  {title} {\bibinfo {title} {Time resolved study of self‐assembled {InAs} quantum dots},\ }\href {https://doi.org/10.1063/1.117827} {\bibfield  {journal} {\bibinfo  {journal} {Applied Physics Letters}\ }\textbf {\bibinfo {volume} {69}},\ \bibinfo {pages} {4087} (\bibinfo {year} {1996})},\ \Eprint {https://arxiv.org/abs/https://pubs.aip.org/aip/apl/article-pdf/69/26/4087/18523780/4087\_1\_online.pdf} {https://pubs.aip.org/aip/apl/article-pdf/69/26/4087/18523780/4087\_1\_online.pdf} \BibitemShut {NoStop}%
\bibitem [{\citenamefont {Fiore}\ \emph {et~al.}(2000)\citenamefont {Fiore}, \citenamefont {Borri}, \citenamefont {Langbein}, \citenamefont {Hvam}, \citenamefont {Oesterle}, \citenamefont {Houdré}, \citenamefont {Stanley},\ and\ \citenamefont {Ilegems}}]{Fiore_2000}%
  \BibitemOpen
  \bibfield  {author} {\bibinfo {author} {\bibfnamefont {A.}~\bibnamefont {Fiore}}, \bibinfo {author} {\bibfnamefont {P.}~\bibnamefont {Borri}}, \bibinfo {author} {\bibfnamefont {W.}~\bibnamefont {Langbein}}, \bibinfo {author} {\bibfnamefont {J.~M.}\ \bibnamefont {Hvam}}, \bibinfo {author} {\bibfnamefont {U.}~\bibnamefont {Oesterle}}, \bibinfo {author} {\bibfnamefont {R.}~\bibnamefont {Houdré}}, \bibinfo {author} {\bibfnamefont {R.~P.}\ \bibnamefont {Stanley}},\ and\ \bibinfo {author} {\bibfnamefont {M.}~\bibnamefont {Ilegems}},\ }\bibfield  {title} {\bibinfo {title} {Time-resolved optical characterization of {InAs/InGaAs} quantum dots emitting at 1.3 $\mu$m},\ }\href {https://doi.org/10.1063/1.126668} {\bibfield  {journal} {\bibinfo  {journal} {Applied Physics Letters}\ }\textbf {\bibinfo {volume} {76}},\ \bibinfo {pages} {3430} (\bibinfo {year} {2000})},\ \Eprint {https://arxiv.org/abs/https://pubs.aip.org/aip/apl/article-pdf/76/23/3430/18549923/3430\_1\_online.pdf}
  {https://pubs.aip.org/aip/apl/article-pdf/76/23/3430/18549923/3430\_1\_online.pdf} \BibitemShut {NoStop}%
\bibitem [{\citenamefont {Wang}\ \emph {et~al.}(1994)\citenamefont {Wang}, \citenamefont {Fafard}, \citenamefont {Leonard}, \citenamefont {Bowers}, \citenamefont {Merz},\ and\ \citenamefont {Petroff}}]{Wang_1994}%
  \BibitemOpen
  \bibfield  {author} {\bibinfo {author} {\bibfnamefont {G.}~\bibnamefont {Wang}}, \bibinfo {author} {\bibfnamefont {S.}~\bibnamefont {Fafard}}, \bibinfo {author} {\bibfnamefont {D.}~\bibnamefont {Leonard}}, \bibinfo {author} {\bibfnamefont {J.~E.}\ \bibnamefont {Bowers}}, \bibinfo {author} {\bibfnamefont {J.~L.}\ \bibnamefont {Merz}},\ and\ \bibinfo {author} {\bibfnamefont {P.~M.}\ \bibnamefont {Petroff}},\ }\bibfield  {title} {\bibinfo {title} {Time‐resolved optical characterization of ingaas/gaas quantum dots},\ }\href {https://doi.org/10.1063/1.111434} {\bibfield  {journal} {\bibinfo  {journal} {Applied Physics Letters}\ }\textbf {\bibinfo {volume} {64}},\ \bibinfo {pages} {2815} (\bibinfo {year} {1994})},\ \Eprint {https://arxiv.org/abs/https://pubs.aip.org/aip/apl/article-pdf/64/21/2815/18503172/2815\_1\_online.pdf} {https://pubs.aip.org/aip/apl/article-pdf/64/21/2815/18503172/2815\_1\_online.pdf} \BibitemShut {NoStop}%
\bibitem [{\citenamefont {Daniels}\ \emph {et~al.}(2013)\citenamefont {Daniels}, \citenamefont {Machnikowski},\ and\ \citenamefont {Kuhn}}]{Daniels_2013}%
  \BibitemOpen
  \bibfield  {author} {\bibinfo {author} {\bibfnamefont {J.~M.}\ \bibnamefont {Daniels}}, \bibinfo {author} {\bibfnamefont {P.}~\bibnamefont {Machnikowski}},\ and\ \bibinfo {author} {\bibfnamefont {T.}~\bibnamefont {Kuhn}},\ }\bibfield  {title} {\bibinfo {title} {Excitons in quantum dot molecules: Coulomb coupling, spin-orbit effects, and phonon-induced line broadening},\ }\href {https://api.semanticscholar.org/CorpusID:118694587} {\bibfield  {journal} {\bibinfo  {journal} {Physical Review B}\ }\textbf {\bibinfo {volume} {88}},\ \bibinfo {pages} {205307} (\bibinfo {year} {2013})}\BibitemShut {NoStop}%
\bibitem [{\citenamefont {Rautert}\ \emph {et~al.}(2019)\citenamefont {Rautert}, \citenamefont {Shamirzaev}, \citenamefont {Nekrasov}, \citenamefont {Yakovlev}, \citenamefont {Klenovsk'y}, \citenamefont {Kusrayev},\ and\ \citenamefont {Bayer}}]{Rautert_2019}%
  \BibitemOpen
  \bibfield  {author} {\bibinfo {author} {\bibfnamefont {J.}~\bibnamefont {Rautert}}, \bibinfo {author} {\bibfnamefont {T.~S.}\ \bibnamefont {Shamirzaev}}, \bibinfo {author} {\bibfnamefont {S.~V.}\ \bibnamefont {Nekrasov}}, \bibinfo {author} {\bibfnamefont {D.~R.}\ \bibnamefont {Yakovlev}}, \bibinfo {author} {\bibfnamefont {P.}~\bibnamefont {Klenovsk'y}}, \bibinfo {author} {\bibfnamefont {Y.~G.}\ \bibnamefont {Kusrayev}},\ and\ \bibinfo {author} {\bibfnamefont {M.}~\bibnamefont {Bayer}},\ }\bibfield  {title} {\bibinfo {title} {Optical orientation and alignment of excitons in direct and indirect band gap (in,al)as/alas quantum dots with type-i band alignment},\ }\href {https://api.semanticscholar.org/CorpusID:102354092} {\bibfield  {journal} {\bibinfo  {journal} {Physical Review B}\ } (\bibinfo {year} {2019})}\BibitemShut {NoStop}%
\bibitem [{\citenamefont {Sivalertporn}\ \emph {et~al.}(2011)\citenamefont {Sivalertporn}, \citenamefont {Mouchliadis}, \citenamefont {Ivanov}, \citenamefont {Philp},\ and\ \citenamefont {Muljarov}}]{Sivalertporn_2011}%
  \BibitemOpen
  \bibfield  {author} {\bibinfo {author} {\bibfnamefont {K.}~\bibnamefont {Sivalertporn}}, \bibinfo {author} {\bibfnamefont {L.}~\bibnamefont {Mouchliadis}}, \bibinfo {author} {\bibfnamefont {A.~L.}\ \bibnamefont {Ivanov}}, \bibinfo {author} {\bibfnamefont {R.}~\bibnamefont {Philp}},\ and\ \bibinfo {author} {\bibfnamefont {E.~A.}\ \bibnamefont {Muljarov}},\ }\bibfield  {title} {\bibinfo {title} {Direct and indirect excitons in semiconductor coupled quantum wells in an applied electric field},\ }\href {https://api.semanticscholar.org/CorpusID:118321953} {\bibfield  {journal} {\bibinfo  {journal} {Physical Review B}\ }\textbf {\bibinfo {volume} {85}},\ \bibinfo {pages} {045207} (\bibinfo {year} {2011})}\BibitemShut {NoStop}%
\bibitem [{\citenamefont {Greve}\ \emph {et~al.}(2011)\citenamefont {Greve}, \citenamefont {McMahon}, \citenamefont {Press}, \citenamefont {Ladd}, \citenamefont {Bisping}, \citenamefont {Schneider}, \citenamefont {Kamp}, \citenamefont {Worschech}, \citenamefont {Hoefling}, \citenamefont {Forchel},\ and\ \citenamefont {Yamamoto}}]{Greve_2011}%
  \BibitemOpen
  \bibfield  {author} {\bibinfo {author} {\bibfnamefont {K.~D.}\ \bibnamefont {Greve}}, \bibinfo {author} {\bibfnamefont {P.~L.}\ \bibnamefont {McMahon}}, \bibinfo {author} {\bibfnamefont {D.}~\bibnamefont {Press}}, \bibinfo {author} {\bibfnamefont {T.~D.}\ \bibnamefont {Ladd}}, \bibinfo {author} {\bibfnamefont {D.}~\bibnamefont {Bisping}}, \bibinfo {author} {\bibfnamefont {C.}~\bibnamefont {Schneider}}, \bibinfo {author} {\bibfnamefont {M.}~\bibnamefont {Kamp}}, \bibinfo {author} {\bibfnamefont {L.}~\bibnamefont {Worschech}}, \bibinfo {author} {\bibfnamefont {S.}~\bibnamefont {Hoefling}}, \bibinfo {author} {\bibfnamefont {A.}~\bibnamefont {Forchel}},\ and\ \bibinfo {author} {\bibfnamefont {Y.}~\bibnamefont {Yamamoto}},\ }\bibfield  {title} {\bibinfo {title} {Ultrafast coherent control and suppressed nuclear feedback of a single quantum dot hole qubit},\ }\href {https://api.semanticscholar.org/CorpusID:118283027} {\bibfield  {journal} {\bibinfo  {journal} {Nature Physics}\ }\textbf {\bibinfo {volume} {7}},\
  \bibinfo {pages} {872} (\bibinfo {year} {2011})}\BibitemShut {NoStop}%
\end{thebibliography}%

\end{document}